\documentclass[a4paper, traditabstract]{aa} 
\pdfoutput=1

\usepackage{graphicx}
\usepackage{subcaption}
\usepackage{caption}
\usepackage{changepage}
\usepackage[usenames,dvipsnames]{color}
\usepackage{xcolor}
\usepackage{booktabs}
\usepackage{txfonts}
\usepackage[bookmarks, colorlinks, breaklinks, ]{hyperref} 
\usepackage{natbib}
\hypersetup{linkcolor=blue,citecolor=MidnightBlue,filecolor=blue,urlcolor=blue} 
\usepackage{hyperref}
\usepackage{textcomp}
\usepackage{bbm}
\usepackage{microtype}
\usepackage{ulem}
\def\lesssim{\mathrel{\hbox{\rlap{\hbox{\lower4pt\hbox{$\sim$}}}\hbox{$<$}}}}
\def\gtrsim{\mathrel{\hbox{\rlap{\hbox{\lower4pt\hbox{$\sim$}}}\hbox{$>$}}}}

\newfont{\gwpfont}{cmssq8 scaled 1000}
\newcommand{\rexcess}{{\gwpfont REXCESS}}
\newcommand{\excpres}{{\gwpfont EXCPReS}}

\usepackage{subcaption}
\usepackage[labelsep=quad,skip=5pt,labelfont=bf]{caption}
\def\msol{{M$_{\odot}$}}

\def\xmm{XMM-{\it Newton}}
\def\planck{{\it Planck}}

\def\M500{M_{500}}
\def\R500{R_{500}}

\def\YX {Y_{\rm X, 500}}

\def\YSZ {Y_{\rm SZ}}

\def\keV {\rm keV}

\def \Mgas  {M_{\rm gas}}

\def \rhoc {\rho_{\rm c}}

\def \rhog {\rho_{\rm gas}}

\def\msol {{\rm M_{\odot}}}

\def\lesssim{\mathrel{\hbox{\rlap{\hbox{\lower4pt\hbox{$\sim$}}}\hbox{$<$}}}}
\def\gtrsim{\mathrel{\hbox{\rlap{\hbox{\lower4pt\hbox{$\sim$}}}\hbox{$>$}}}}

\newcommand{\propsim}{\lower 3pt \hbox{$\, \buildrel {\textstyle
       \propto}\over {\textstyle \sim}\,$}}

\usepackage[dvipsnames]{xcolor}
\usepackage{multirow}

\begin{document}
\title{Galaxy cluster thermodynamic profiles beyond self-similarity}

\author{E. Baraldi\inst{1}, G.W. Pratt\inst{1} \and E. Pointecouteau \inst{2}}
\authorrunning{E. Baraldi et al.}
\titlerunning{Investigating deviations from self-similarity in a sample of massive galaxy clusters}
 \institute{
Université Paris Cité, Université Paris-Saclay, CEA, CNRS, AIM de Paris-Saclay, 91191 Gif-sur-Yvette, France \\
\email{edoardo.baraldi@cea.fr}
 \and
 IRAP, CNRS, Université de Toulouse, CNES, UT3-UPS, (Toulouse), France\\
}
\date{Accepted 19 October 2026}
\abstract{
Galaxy cluster formation is, to first order, a self-similar and scale-free process of gravitational collapse. However, the thermodynamic properties of the intracluster medium often deviate from self-similar predictions. We investigate these deviations using a sample of 92 galaxy clusters selected through their Sunyaev-Zeldovich effect signal, spanning a mass range of $M_{500c}=[0.5-20] \times 10^{14}\textup{M}_{\odot}$, and with redshifts $0.08 < z < 1.13$. 
Adopting a flexible framework that allows for radial variations in both redshift and mass scaling, we characterise the behaviour of scaled density, temperature, pressure, and entropy profiles, while simultaneously constraining their intrinsic scatter. When scaled according to the critical density of the Universe, we detect significant radial deviations from self-similar expectations in both redshift and mass scalings in all profiles. These deviations are strongest in the core regions, where baryonic processes dominate, but remain measurable out to large radii ($\sim R_{500c}$), indicating that even cluster outer regions are not fully described by purely gravitational evolution. We investigate the impact of an offset, a mass-dependence, and increased scatter between the mass proxy and the true mass on the radial mass scaling, finding that while its shape remains stable, its normalisation depends on the latter two effects. The pressure is the only thermodynamic quantity that exhibits approximately self-similar behaviour: outside the core it evolves self-similarly, with no dependence on mass at the $2\sigma$ level. In contrast, the entropy exhibits the strongest deviations from self-similarity as a function of radius. 
For all profiles, the intrinsic scatter peaks in the central regions and reaches a minimum 
at intermediate radii ($\lesssim 20 \%$ at $\sim 0.2$--$0.8 \ R_{500c}$), making this range optimal to reduce scatter in scaling relations. 
We conclude that the thermodynamic distributions exhibit non-self similar, mass-dependent scaling extending well beyond the core regime, likely shaped by the interplay between baryonic processes and ongoing accretion across all radii.}


\maketitle

\section{Introduction}\label{sec:intro}
In the $\Lambda$ cold dark matter scenario ($\Lambda$CDM), the hierarchical gravitational collapse of dark matter halos drives the structure formation from the last massive to galaxy clusters. The rarefied gas of the intra-cluster medium (ICM) follows this collapse, heating through shocks and compression to temperatures up to $10^7 -10^8$ K, becoming a fully ionised plasma, emitting in the X-ray band through bremsstrahlung radiation and line emissions. Photons from the cosmic microwave background (CMB) interact with the hot ICM plasma through inverse Compton scattering, producing a spectral shift of the CMB emission in the direction of clusters, the Sunyaev-Zeldovich effect \citep[SZE;][]{sunyaev_observations_1972}.

If the hierarchical collapse model is completely driven by gravity, the process is self-similar and scale-free \citep[][]{kaiser_evolution_1986, bryan_statistical_1998}. With the additional assumption of a constant gas mass fraction, the various integrated properties of the ICM display power-law relationships as a function of the total cluster mass. Deviations from this ideal power-law scaling arise essentially from non-gravitational processes - primarily feedback from active galactic nucleus (AGN), star formation, stellar feedback, and radiative cooling of the gas -- and have been predicted \citep[e.g.][]{voit_tracing_2005, planelles_role_2014, ettori_icmz_2023} and observed both in integrated thermodynamic properties \citep[e.g.][]{maughan_lx-yx_2007,croston_galaxy-cluster_2008, pratt_galaxy_2009, giodini_scaling_2013, mantz2018, bulbul2019, lovisari_scaling_2022} and in the associated radial profiles \citep[e.g.][]{arnaud_universal_2010, pratt_gas_2010, maughan_self-similar_2012,mantz2016, mcdonald_remarkable_2017,pra22, sotira_impact_2025, oppenheimer_introducing_2025}. 

The advent of SZE surveys \citep[][]{hasselfield_atacama_2013, bleem_galaxy_2015, collaboration_planck_2016, bocquet_cluster_2019, zubeldia_planck_2024, cherouvrier_nika2_2025} has allowed us to explore the galaxy cluster population at high redshift, and to study how this population changes over time. Due to the differing degrees of correlation between the X-ray and SZE signal properties and the underlying cluster mass, SZE and X-ray selections sample the underlying population in distinct ways. It is likely that SZE selected clusters, which tend to be less X-ray luminous at a given mass than X-ray selected clusters, exhibit a somewhat different distribution of dynamical states than X-ray selected systems \citep[e.g.][]{collaboration_planck_2014, rossetti_measuring_2016, bartalucci_most_2019}. 
Indeed, the dynamical state may be thought of as the third fundamental cluster property, after the mass and the redshift.

Deviations from self-similarity can give important clues about the effects of non-gravitational processes on the thermodynamic properties of the cluster population. In this paper, we use radial profiles of the gas density and temperature of 92 objects selected via their SZE signal to examine how scaled cluster properties deviate from self-similarity in concentric radial shells. Quantifying the radial dependence of the mass and redshift scaling factors across the sample, we find the strongest deviations from self-similar behaviour in the core regions, where the effects of non-gravitational processes are the most pronounced: unsurprisingly, cluster cores do not evolve in a self-similar fashion. We show that these deviations can persist up to a large radius and that this persistence is mass-dependent. We then derive the universal scaled thermodynamic profiles of the sample and the associated intrinsic scatter, using the posterior distributions of the radial dependence of the mass and redshift scaling factors as additional priors. We compare the resulting intrinsic scatter profiles to those from previous work \citep{ghirardini_universal_2019,pra22,riva_chex-mate_2024,munoz-echeverria_chex-mate_2025}, finding that our intrinsic scatter is generally somewhat smaller outside the core regions. We investigate the extent to which a different cluster selection (X-ray, SZE) or dynamical state (relaxed, disturbed) affects our results. Finally, we quantify the impact of the absolute mass scale and the scatter about the various mass proxy relations on our results. 

We assume a flat $\Lambda$CDM cosmology with $\Omega_m=0.3$, $\Omega_{\Lambda}=0.7$, and $H_0=70$ km s$^{-1}$ Mpc$^{-1}$. All logarithmic quantities, including intrinsic scatters, are expressed in base $e$ (unless otherwise specified), and the quoted uncertainties correspond to the $68\%$ confidence interval. 

\section{Dataset and analysis}\label{sec:Analysis}

\subsection{Datasets}\label{subsec:SZDataset}
The dataset consists of 92 SZE-selected clusters at $0.08<z<1.13$ \citep[][]{pra22}. The sample comprises 44 objects at $z<0.5$ from the \textit{Planck} Early SZ sample \citep[Planck ESZ,][]{collaboration_planck_2011}, which \cite{bartalucci_most_2019} show is representative of the full ESZ, and 48 objects observed by XMM-\textit{Newton} in a series of three large programmes as part of the M2C project. 
The sample spans a total mass range of $M_{500c}=[0.5-20] \times 10^{14} \textup{M}_{\odot}$ and lies at redshifts of $0.08<z<1.13$. A full description of this sample is given in the appendix of \cite{pra22}. Figure~\ref{fig:sample} shows the distribution of the sample in the redshift-mass plane, along with relevant comparison samples. 

\begin{figure}
\includegraphics[width=8cm]{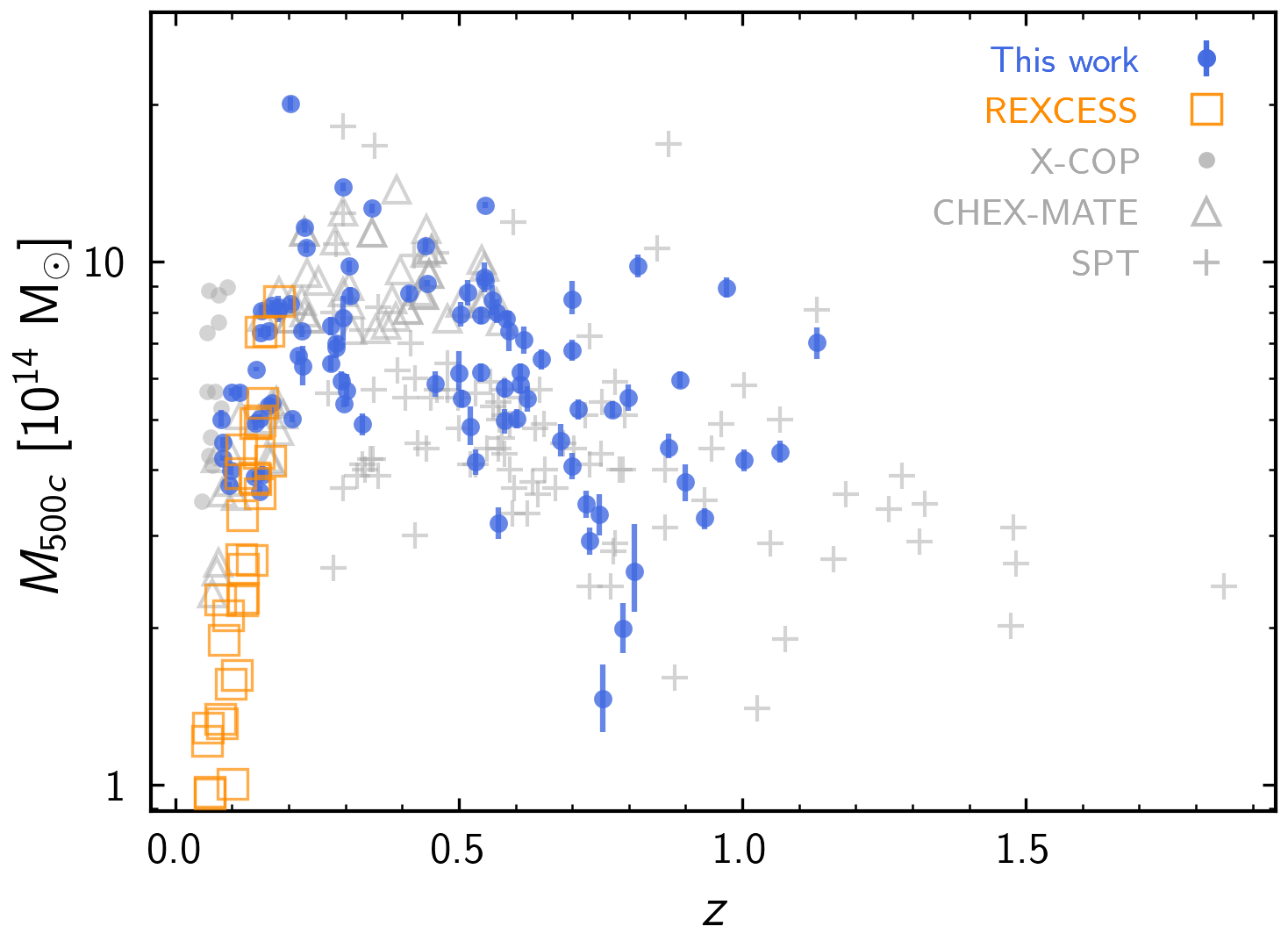} \\ 
\caption{\footnotesize Redshift-mass distribution of the clusters considered in this work compared to other notable samples (grey points). The SZE-selected clusters (blue points) comprise a subset of 44 systems from the \planck\ Early SZ sample \citep{collaboration_planck_2011} at $z<0.5$ and a further 48 clusters at $z>0.5$. \rexcess \ \citep[orange open squares]{boehringer_representative_2007} is a X-ray-selected samples of 31 systems at $z<0.25$. The other comparison samples in grey are \ref{sec:discussion}: X-COP \citep{ghirardini_universal_2019}, CHEX-MATE \citep{riva_chex-mate_2024, munoz-echeverria_chex-mate_2025}, and SPT \citep{mcdonald_remarkable_2017}.}
\label{fig:sample}
\end{figure}

We also considered the X-ray selected sample \rexcess \ \citep[][]{boehringer_representative_2007, croston_galaxy-cluster_2008, pratt_gas_2010, arnaud_universal_2010}, shown in Fig.~\ref{fig:sample}. This sample lies at redshift of $0.055<z<0.183$, and includes clusters with X-ray luminosities above $0.4 \times10^{44}$ erg s$^{-1}$. It spans a narrower mass range of $10^{14}\,\textup{M}_{\odot}<M_{500c}<10^{15}\,\textup{M}_{\odot}$ and does not include the most massive systems. The X-ray and SZE-selected samples have six systems in common. 

The data reduction was performed following the procedures described in \cite{croston_galaxy-cluster_2008,pratt_galaxy_2009, pratt_gas_2010, pra22}. The event lists were reprocessed with XMM-SAS and the associated calibrated files produced. Standard filtering and \textsc{Pattern}-selection were applied for clean events for which soft proton flares were removed. The instrumental and particle background were obtained from custom stacked data files derived from observations obtained with filter wheel in the CLOSED position (FWC), renormalised using the count rate in a high energy band free of cluster emission. Point sources were detected with a wavelet algorithm and excised. 

\subsection{Thermodynamic profiles}\label{subsec:thermprof}

Surface brightness profiles, extracted in annular bins centred on the X-ray peak, were then produced from the point source-excised, vignetting-corrected event files in the $[0.3-2]$ keV band, and background-subtracted. The procedure used to obtain the deprojected, point spread function (PSF)-corrected density profiles is described in \cite{croston_improved_2006}. In summary, the vignetting-corrected background-subtracted $[0.3-2]$ keV surface brightness profiles were converted to gas density by calculating an emissivity profile $\Lambda (T, Z)$ in XSPEC \citep[][]{arnaud_xspec_1996}, taking into account the absorption and instrumental response, and using a model description for the projected temperature profile and abundance profile \citep[for further details see e.g.][]{pratt_entropy_2003, croston_improved_2006, croston_galaxy-cluster_2008, bartalucci_resolving_2017}. The electron density measured in X-ray, $n_{\textup{e}}$, was converted to gas density, $\rho_{\textup{gas}}=n_{\textup{e}} \times (\mu_{\textup{e}} m_{\textup{p}})$, where $m_{\textup{p}}$ is the proton mass and $\mu_{\textup{e}}=1.14$ is the mean molecular weight per free electron:
\begin{equation}
\label{eqn:rhogas}
\rhog=1.92 \times 10^{-24} \left( \frac{n_{\textup{e}}}{\rm cm^{-3}} \right) \ \rm g \ \rm cm^{-3}.
\end{equation}
The projected 2D temperature profiles were derived in logarithmically spaced annular bins centred on the peak of the X-ray emission following the procedures described in  \cite{pratt_gas_2010}. The FWC-subtracted spectrum in the annular region external to the cluster emission was described with a model consisting of two thermal 
components plus an absorbed power law with a fixed slope of $\Gamma=1.4$. This best-fitting background model, renormalised appropriately to the ratio of the surface area of the extraction region, was added as an extra component in each annular. In the annular cluster emission region, the model was then fitted, keeping the absorption fixed to the HI value \citep[][]{kalberla_leidenargentinebonn_2005}, letting the temperature and metallicity free to vary. The 3D temperature profiles were reconstructed by fitting the projected and PSF-corrected parametric model introduced in \cite{vikhlinin_chandra_2006}, weighted for the bias introduced by fitting an isothermal model to multi-temperature plasma emission using the scheme proposed by \cite{vikhlinin_chandra_2006}, to the observed 2D annular temperature profile. We recomputed the best-fitting parametric 3D temperature profiles on the same grid as that of the deprojected, deconvolved density profiles \citep[][]{arnaud_universal_2010, pratt_gas_2010}, and computed the pressure and entropy profiles following the astrophysical convention, $P(r)=k_{\rm{B}}T(r)\,n_{\textup{e}}(r)$ and $K(r)=k_{\rm{B}}T(r)/n_{\textup{e}}^{2/3}$ \citep[][]{ponman_imprint_1999, lloyd-davies_entropy_2000}, respectively, with $k_{\rm{B}}$ being the Boltzmann constant.

\section{Scaling model}\label{subsec:selfsimilarmodel}
\subsection{Basic self-similar formalism}\label{subsec:selfsimilar}

In the self-similar model \citep[e.g.][]{kaiser_evolution_1986}, galaxy clusters can be fully described by their redshift, $z$, and mass, $M_\Delta=\ \Delta \rho_{\textup{ref}}(z)(4\pi /3) R_{\Delta}^3$, and cluster profiles rescaled by their corresponding virial quantities should follow a universal shape. These virial quantities can be defined by a fixed overdensity with respect to the critical density, $\rhoc(z) = 3H_0^2\, E^2(z)/8 \pi G$, or the mean density, $\rho_m(z) = \rho_c(z)\, \Omega_m(z)$, where $E(z)=[\Omega_{\rm m}(1+z)^3+\Omega_{\rm \Delta}]^{1/2}$ is the evolution of the Hubble parameter with redshift in a flat cosmology. Numerical simulations \citep{lau2015} showed that the inner regions of galaxy clusters typically covered by X-ray observations ($R \lesssim R_{500c}$) follow a more self-similar behaviour when the cluster radial profiles are rescaled with respect to $\rhoc(z)$. Conversely, \citeauthor{lau2015} found that scaling by $\rho_m(z)$ was better suited to describing the structure of galaxy clusters in the outer regions.
Since our thermodynamic profiles are derived from X-ray observations, and therefore mainly probe the inner regions of galaxy clusters, we chose to rescale with respect to $\rhoc(z)$.  
We investigate scaling by $\rho_m(z)$ in Appendix~\ref{app:mean_matt_dens}.

Total masses, $M_{500c}$, and the corresponding radii, $R_{500c}$, were estimated from the mass proxy $Y_{\textup{X},500c}=\Mgas\times T$, using the calibration in \citet[][]{arnaud_universal_2010}.
The quantity $\YX$$_{\rm c}$ is the X-ray analogue of the integrated SZ signal $\YSZ$, which has been shown to be a low scatter mass proxy in the numerical simulations of \cite{kravtsov_new_2006}, irrespective of dynamical state. This mass proxy is appropriate since the present cluster sample contains objects in different dynamical states, for which the hydrostatic equilibrium assumption does not necessarily hold.

\begin{figure*}
\centering
\includegraphics[width=16cm]{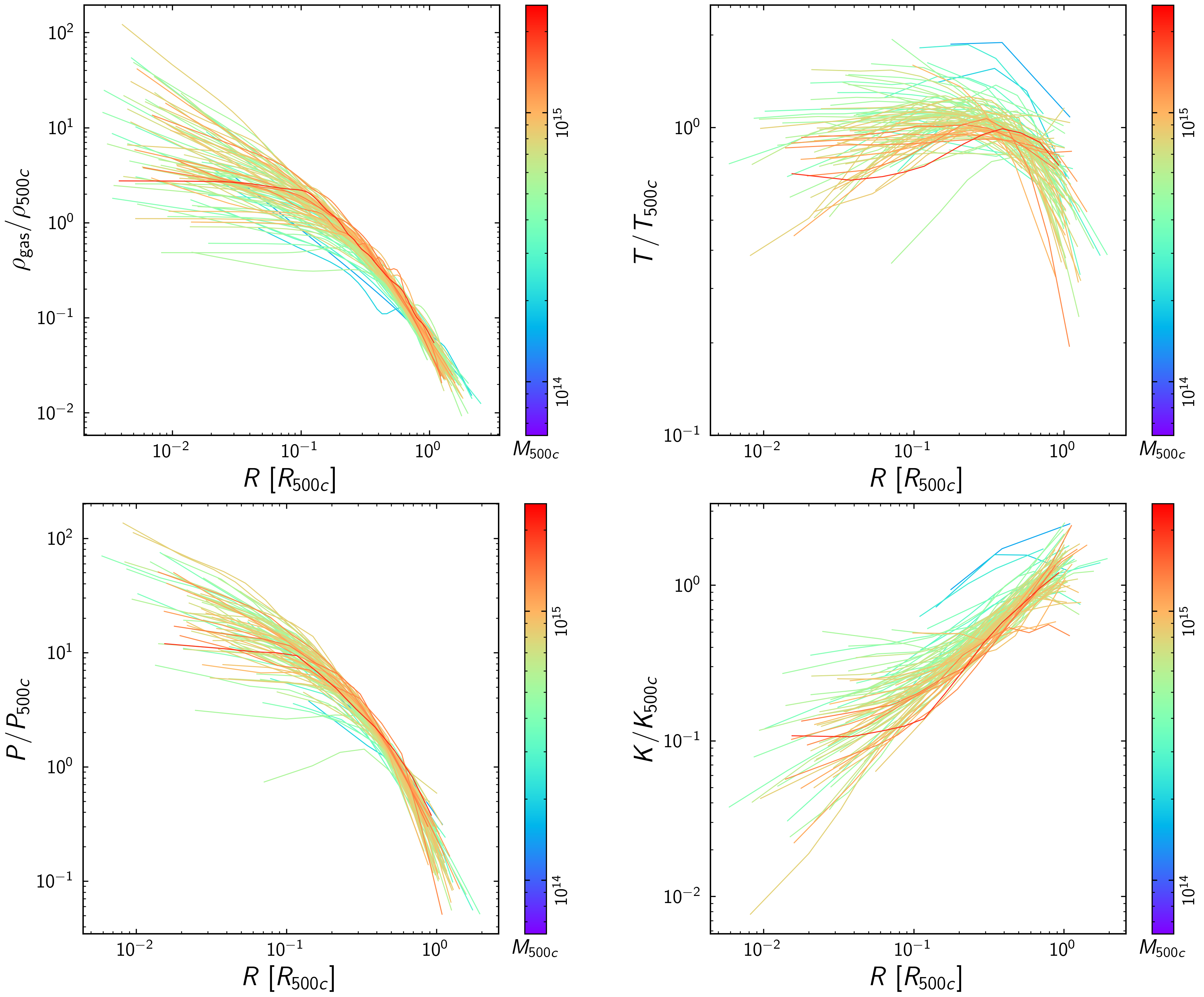}
\caption{\footnotesize Density (top left panel), temperature (top right panel), pressure (bottom left panel), and entropy (bottom right panel) profiles for the 92 SZE-selected clusters, normalised by $\rho_{500}$, $T_{500}$, $P_{500}$, and $K_{500}$, respectively. The radial bins are scaled by $\R500$ and the profiles are colour-coded by mass $M_{500}$, computed by total mass proxy $\YX$ \citep[][]{kravtsov_new_2006}, using the scaling calibrated in \citep[][]{arnaud_universal_2010}. }
\label{fig:thermoprofiles}
\end{figure*}

Within this framework, the gas density was rescaled by
\begin{equation}
\label{eqn:rho500}
\rho_{500c}=500 \ \rhoc(z)=4.603\times10^{-27} E(z)^{2} \ \textup{g} \ \textup{cm}^{-3}.
\end{equation}
The other thermodynamic quantities were rescaled by \citep[e.g.][]{ghirardini_universal_2019}
\begin{equation}\begin{split}
\label{eqn:T_P_K_500}
T_{500c}&=5.58 \ \left( \frac{M_{500c}}{5\times10^{14}\, \msol} \right)^{2/3} \ E(z)^{2/3} \ \rm \keV, \\
P_{500c}&=2.16 \times 10^{-3} \ \left( \frac{M_{500c}}{5\times10^{14}\, \msol} \right)^{2/3} \ E(z)^{8/3} \ \rm \keV \ \rm cm^{-3}, \\
K_{500c}&=1050 \ \left( \frac{M_{500c}}{5\times10^{14}\, \msol} \right)^{2/3} \ E(z)^{-2/3} \ \rm \keV \ \rm cm^{2},
\end{split}
\end{equation}
where we assumed a pivot mass of $M_p = 5\times10^{14}\, \msol$, a universal gas fraction of $f_b = \Omega_b/\Omega_m = 0.16$ \citep{planck_cosmo}, a mean molecular weight per particle of $\mu = 0.6$, and a mean molecular weight per electron of $\mu_e=1.14$ \citep{and89}. 
Figure \ref{fig:thermoprofiles} shows the scaled density, temperature, pressure, and entropy profiles for the 92 SZE-selected objects, colour-coded by mass. Scatter about the universal shape is driven by the variation in the dynamical state across the population, and, in the central regions, by the effect of non-gravitational processes. Such processes can include radiative cooling, AGN feedback, and star formation \citep{kaiser_evolution_1986}, alongside non-thermal pressure support generated by turbulence \citep{leau2009, eckert2019, ettorieckert2022, chappuis_chex-mate_2025}. The colour-coding by mass further highlights that higher-mass clusters scale differently from lower-mass systems. These trends suggest an additional dependence of the scaling on $\M500$ and/or $E(z)$ with respect to those predicted by a self-similar model. 
In \cite{pra22} and \cite{riva_chex-mate_2024}, modifications to the self-similar model were tested by introducing additional mass and redshift dependencies in the scaling relations. The observed radial dependence of the profile dispersion --shifting from core regions dominated by baryon physics to outer radii that more closely follow self-similar expectations (see Figure \ref{fig:thermoprofiles}) --motivates a more flexible modelling approach. We therefore extended our analysis to include radially varying mass and evolution scaling factors. 

\subsection{Modified scaling models}

\subsubsection{Modified piecewise linear model}\label{subsec:modeling}

We first used a piecewise power law fitting approach. Working with scaled radii $x=R/R_{500c}$, we described the data in a series of fixed scaled radial bins using a model that allows for more flexibility for the mass and redshift scaling factors. In each bin of scaled radius, the data, scaled by the self-similar expectations, were fitted with a power law model given by
\begin{equation}
\label{eqn:QQ500}
\frac{Q}{Q_{500c}} = A \cdot \left( \frac{M_{500c}}{5\times10^{14} \ \msol} \right)^{\delta\alpha_M} \cdot
E(z)^{\delta\alpha_{\rm z}} \cdot \rm exp[{\pm\sigma}],
\end{equation}
where $Q/Q_{500}$ is the scaled thermodynamic quantity in question, and $A$, $\delta\alpha_M$, and $\delta\alpha_z$ are the normalisation and the mass and redshift dependencies relative to the self-similar predictions, respectively. In this framework, $\delta\alpha_z$ and $\delta\alpha_M$ are identically zero for all quantities if clusters are fully self-similar. The total variance, $\sigma^2$, is defined as 
\begin{equation}
\label{eqn:sigma}
\sigma^2 = \sqrt{\sigma_{\rm stat}^2 + \sigma_{\rm int}^2},
\end{equation}
where $\sigma_{\rm stat}$ and $\sigma_{\rm int}$ are the statistical and intrinsic scatter, respectively, evaluated on the natural logarithm of the quantities in question. The necessity of accounting for intrinsic scatter is well established in the literature \citep[e.g. ][]{sayers2013,mantz2016, ghirardini_universal_2019, sayers_evolution_2023}. This scatter quantifies the inherent diversity within the cluster population, arising from both the range of dynamical states and the impact of non-gravitational processes on the baryonic physics. 

\subsubsection{Modified models for scaled thermodynamic profiles}\label{subsec:analytmodel}

\paragraph{Generic approach}

We described the profiles, scaled for self-similar expectations, with a model consisting of an analytical functional form, a normalisation that varies with radius, mass and redshift, and a radial variance. The model is described as follows:
\begin{equation}
\label{eqn:medianprofile}
\frac{Q(x, z, M_{500c})}{Q_{500c}}= N(x, z, M_{500c}) \cdot f(x) \cdot  {\rm exp}[\pm\sigma(x)],
\end{equation}
where $x=R/R_{500c}$, $Q/Q_{500c}$ is the rescaled thermodynamic quantity, and $f(x)$ is the functional form for the thermodynamic quantity, $Q$, in question. The normalisation, $N(x, z, M_{500c})$, describes deviations from standard self-similarity in terms of the radial dependence of both mass and redshift scalings as mentioned in Sect.~\ref{subsec:modeling} as
\begin{equation}
\label{eqn:normmassred}
N(x, z, M_{500c}) = \left( \frac{M_{500c}}{5\times10^{14} \ \msol} \right)^{\delta\alpha_M(x)} \cdot
E(z)^{\delta\alpha_z(x)},
\end{equation}
in which we assumed the same power-law dependence on $M_{500}$ and $E(z)$ as given in Eq.~\ref{eqn:QQ500}, while adopting the posteriors of the previously determined slope values as priors in the fitting procedure. The variance term, $\sigma^2(x)$, is defined as the quadratic sum of the 
statistical error, $\sigma_{\rm stat}$, and the intrinsic scatter term, $\sigma_{\rm int}$, in the natural logarithm of the measured thermodynamic quantity, which was allowed to vary according to a quadratic functional form \citep[][]{ghirardini_universal_2019},
\begin{equation}
\label{eqn:intrscatter}
\sigma_{\rm int}(x) = \sigma_1 \ {\rm ln}^{2} \left(\frac{x}{x_0}\right) \ + \ \sigma_0,
\end{equation}
with $\sigma_1$ the width of the natural log-parabola, and $\sigma_1$ and $\sigma_0$ the intercept and the location of the minimum of the natural log-parabola, respectively. 

\paragraph{Density model}\label{subsec:densmodel}
We described the scaled density profiles with a generalised Navarro-Frenk-White (gNFW) model \citep[e.g.][]{nagai_effects_2007},
\begin{equation}
\label{eqn:GNFW}
f(x)=\frac{n_0}{(x/x_{\rm s})^{\alpha}[1+(x/x_{\rm s})^{\gamma}]^{(3\beta-\alpha)/\gamma}},
\end{equation}
where $n_0$ is the normalisation, $x_{\rm s}$ is the scaling radius, and the parameters $(\alpha, \gamma, \beta)$ are the central ($x\ll x_{\rm s}$), intermediate ($x\sim x_{\rm s}$), and outer ($x\gg x_{\rm s}$) slopes, respectively. The case in which $\alpha=0$ and $\gamma=2$ corresponds to the standard $\beta$ model \citep[][]{cavaliere_x-rays_1976}, while the case $\gamma=2$ corresponds to the AB model introduced by \citet[][]{pratt_mass_2002}. 

\paragraph{Temperature model}\label{subsec:tempmodel}
The deprojected temperature profiles were fitted with the functional form described in \cite{vikhlinin_chandra_2006}, which provides an accurate description of the temperature profiles across both central and outer cluster regions,
\begin{equation}
\label{eqn:tempmodl}
f(x)=T_0\frac{(T_{\rm min}/T_0)+(x/x_{\rm cool})^{\alpha_{\rm cool}}}{1+(x/x_{\rm cool})^{\alpha_{\rm cool}}} \frac{1}{(1+(x/x_{\rm t})^{2})^{c/2}},
\end{equation}
where ($T_{\rm min},r_{\rm cool},\alpha_{\rm cool}$) is the set of parameters describing the temperature decline in the central regions, while ($r_{\rm t}, c$) are used to represent the temperature profiles outside the central cooling region; $T_0$ is the central temperature.

\paragraph{Pressure model}\label{subsec:pressmodel}
The pressure profiles were described with the gNFW model introduced by \citet{nagai_effects_2007}:
\begin{equation}
\label{eqn:GNFWpress}
f(x)=\frac{P_0}{(c_{500c}x)^{\alpha}[1+(c_{500c}x)^{\gamma}]^{(\beta-\alpha)/\gamma}},
\end{equation}
where $P_0$ is the normalisation and central pressure, the slopes $(\alpha, \gamma, \beta)$ represent the inner, intermediate, and outer slopes, respectively, and $c_{500c}=R_{500c}/x_{\rm s}$, with $x_{\rm s}$ as the scaling radius.

\paragraph{Entropy model}\label{subsec:entropymodel}
The entropy profiles were fitted throughout the entire radial range with a power law with a constant entropy floor \citep[][]{cavagnolo_intracluster_2009}:
\begin{equation}
\label{eqn:powerlaw_entr}
f(x)=K_0 \ + \ K_1 \ \cdot \ x^{\alpha},
\end{equation}
where $K_0$ is the core entropy, $K_1$ is the normalisation, and $\alpha$ is the power law index. 

\begin{figure*}[!ht]
\centering
\includegraphics[width=16cm]{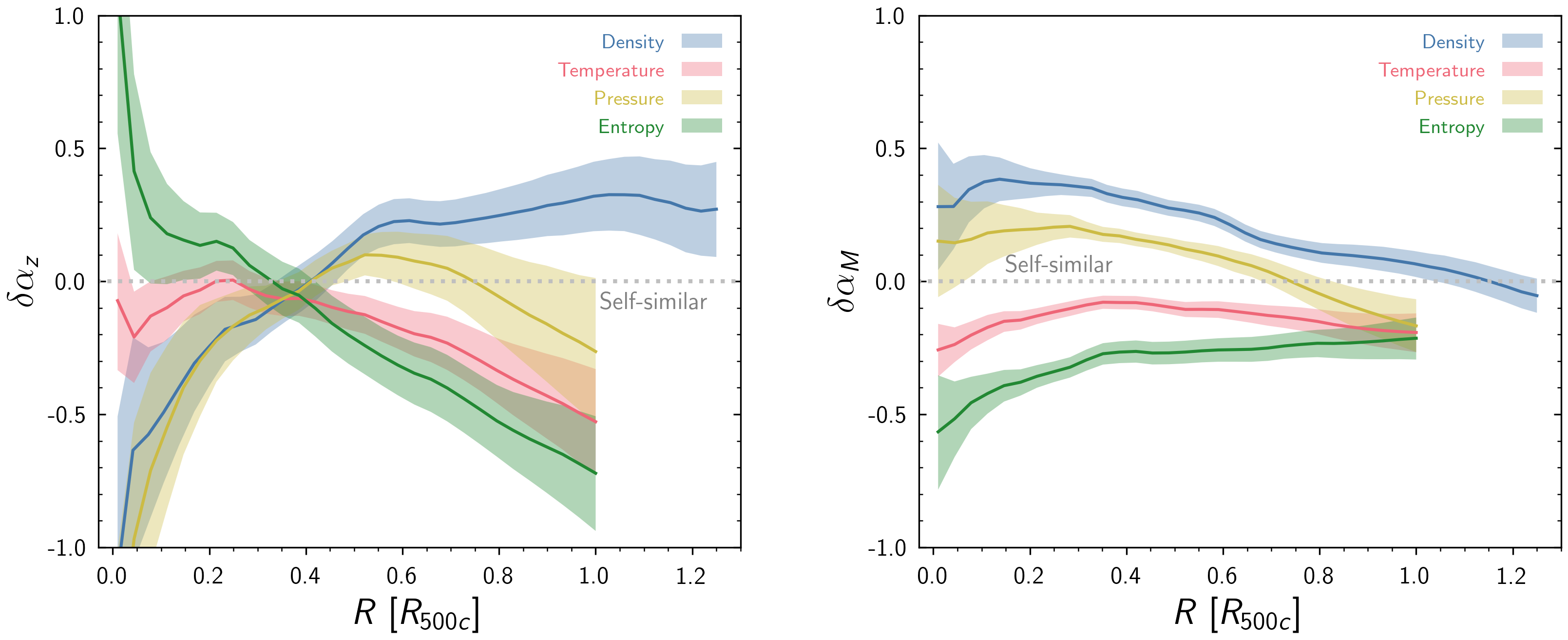} \\ 
\caption{\footnotesize Radial departures from self-similar scaling of the redshift (left) and mass (right) scaling factors, obtained for the density (blue), temperature (red), pressure (yellow), and entropy (green) following the piecewise power-law fitting procedure described in Section \ref{subsec:modeling}. The dotted grey line represents the expected values from a self-similar model. The shaded areas indicate the $1\sigma$ uncertainties of the best-fitting values.}
\label{fig:alpha_M_z}
\end{figure*}

\section{Fitting procedure}\label{sec:fitting_procedure}

In our piecewise power-law approach (Sect.~\ref{subsec:modeling}) we defined 40 linearly spaced radial bins in the $[0.01-1.25]$ $\R500$ radial range for the density, and 30 linearly spaced radial bins in the $[0.01-1.0]$ $\R500$ radial range for the temperature, pressure, and entropy. Where clusters lacked radial coverage, the thermodynamic profiles were interpolated and/or extrapolated to match the boundaries of the adopted radial regime (see Appendix \ref{app:extrapol_prof}). The model (Eq. \ref{eqn:QQ500}) assumes that the scaled thermodynamic quantities within each radial bin are independently described by Gaussian distributions, $Q_{\textup{data},i}/Q_{500c,i}$. Consequently the joint likelihood is defined as the product of these individual distributions,
\begin{equation}
\label{eqn:likelihood_prof_fitting}
\mathcal{L} = \prod_{i} \frac{1}{\sqrt{2\pi}\sigma_i} \textup{exp} \left[-\frac{(Q_{\textup{data},i}/Q_{500c,i}-Q_{\textup{mod},i})^2}{2\sigma^2}\right],
\end{equation}
with $Q_{\textup{mod},i}$ corresponding to the modelled thermodynamic quantity in question. In the scaled thermodynamic profile approach, the data are modelled following Eq. \eqref{eqn:medianprofile}. Consequently, in the joint likelihood function (Eq. \ref{eqn:likelihood_prof_fitting}), $Q_{\textup{mod},i}$ represents the thermodynamic profile models, while the terms  $Q_{\textup{data},i}/Q_{500c,i}$ denote each individual scaled thermodynamic profile.

We used Markov chain Monte Carlo (MCMC) sampling to fit the free parameters. Given the large number of free parameters, we relied on the {\tt jax}-based code \citep{jax2018github} to enable the automatic differentiation of distributions, and to perform a fast sampling of the posterior probability distribution. To simplify the optimisation process, we performed the fitting in natural logarithmic space. Bayesian inference was undertaken using the No-U-Turn sampler \citep[NUTS,][]{hoffman_no-u-turn_2011} as implemented in the {\tt numpyro} package \citep[][]{bingham_pyro_2018, phan_composable_2019}. For the piecewise approach, we executed four chains of 8000 steps each, following a tuning phase of 4000 steps. When fitting the profiles, we used four chains of 30000 steps each, with a tuning of 15000 steps. We considered the MCMC chains converged if the $\hat{R}<1.01$ condition was satisfied \citep[][]{gelman_inference_1992, vehtari_rank-normalization_2021}. 

The mass estimates -- from which $R_{500c}$ is derived and which are used to scale both the radius ($x=r/R_{500c}$) and the thermodynamical quantities ($Q_{500c}$) -- are subject to uncertainties.
We accounted for these by implementing a Bayesian hierarchical model wherein cluster masses were sampled according to Gaussian priors centred on their observed values, thereby modifying the likelihood to include the mass sampling. The posterior distributions of the parameters of interest were obtained by marginalising over the mass parameters, after assessing that there were no significant offsets between the sampled masses and the initial estimates (see Section~\ref{sec:results}). The priors adopted for the piecewise and profile fitting approaches are summarised in Tables~\ref{table:priors} and~\ref{table:prior_and_best_fit}, respectively.

\begin{table}
\caption{\footnotesize Mass and redshift scaling priors.}
\centering                                      
\begin{tabular}{c c}          
\hline\hline                        
Parameters & Priors  \\   
\hline                                   
$A$ & $\mathcal{U}(-5, 5)$ \\
$\alpha_{\rm M}$ & $\mathcal{U}(-2, 2)$ \\
$\alpha_{\rm z}$ & $\mathcal{U}(-2, 2)$ \\
$\sigma$ & $\mathcal{HN}(0, 1^2)$ \\
$\textup{ln}(M_{500c,i})$ & $\mathcal{N}(\textup{ln}(M^{\rm cat}_{500c,i}), \sigma_{M_{500c,i}}^2)$ \\
\hline                                             
\end{tabular}
\tablefoot{We indicate Uniform priors with $\mathcal{U}(a, b)$. We assumed Uniform priors for all the parameters, except for $\sigma$, for which we adopted a half-Gaussian distribution ($\mathcal{HN}$) with variance equal to $1$. For each cluster $i$, we assumed a natural log-normal distribution for the mass, based on $M^{\rm cat}_{500c,i}$ values reported in the catalogues.}
\label{table:priors}      
\end{table}

\section{Results}\label{sec:results}

\begin{table*}
\caption{\footnotesize Best-fitting deviations of the redshift and mass slopes, $\alpha_z (R)$ and $\alpha_M (R)$, relative to the self-similar predictions for density, temperature, pressure, and entropy profiles, evaluated in radial bins of $\Delta R = 0.2\,R_{500c}$.} \label{table:mass_scale_best}     
\centering                                     
\begin{tabular}{c c c c c c c c}         
\toprule
\toprule
& & {[$0-0.2$]} $R_{500c}$ & {[$0.2-0.4$]} $R_{500c}$ & {[$0.4-0.6$]} $R_{500c}$ & {[$0.6-0.8$]} $R_{500c}$ & {[$0.8-1$]} $R_{500c}$ & {[$1-1.25$]} 
$R_{500{c}}$\\
\midrule 
\multirow{2}{*}{Density} & $\delta \alpha_z$ & $-0.7 \ \pm \ 0.3$ & $-0.2 \ \pm \ 0.1$ & $0.14 \ \pm \ 0.08$ & $0.24 \ \pm \ 0.09$ & $0.27 \ \pm \ 0.11$ &
$0.28 \ \pm \ 0.14$ \\
& $\delta \alpha_M$ & $0.34 \ \pm \ 0.10$ & $0.37 \ \pm \ 0.04$ & $0.28 \ \pm \ 0.03$ & $0.15 \ \pm \ 0.03$ & $0.09 \ \pm \ 0.04$
& $0.01 \ \pm \ 0.05$ \\
\midrule
\multirow{2}{*}{Temperature} & $\delta \alpha_z$ & $-0.10 \ \pm \ 0.14$ & $0.01 \ \pm \ 0.07$ & $-0.09 \ \pm \ 0.07$ & $-0.22 \ \pm \ 0.10$ & $-0.43 \ \pm \ 0.16$ & / \\
& $\delta \alpha_M$ & $-0.21 \ \pm \ 0.05$ & $-0.13 \ \pm \ 0.03$ & $-0.11 \ \pm \ 0.03$ & $-0.14 \ \pm \ 0.04$ & 
$-0.17 \ \pm \ 0.06$ & / \\ 
\midrule
\multirow{2}{*}{Pressure} & $\delta \alpha_z$ & $-0.97 \ \pm \ 0.35$ & $-0.11 \ \pm \ 0.12$ & $0.02 \ \pm \ 0.09$ & $0.01 \ \pm \ 0.14$ & $-0.20 \ \pm \ 0.23$ & 
/ \\
& $\delta \alpha_M$ & $0.17 \ \pm \ 0.14$ & $0.23 \ \pm \ 0.04$ & $0.11 \ \pm \ 0.03$ & $0.01 \ \pm \ 0.05$ & 
$-0.11 \ \pm \ 0.09$ & / \\ 
\midrule
\multirow{2}{*}{Entropy} & $\delta \alpha_z$ & $0.48 \ \pm \ 0.24$ & $0.09 \ \pm \ 0.11$ & $-0.16 \ \pm \ 0.11$ & $-0.37 \ \pm \ 0.12$ & $-0.60 \ \pm \ 0.17$ & / \\
 & $\delta \alpha_M$ & $-0.48 \ \pm \ 0.09$ & $-0.38 \ \pm \ 0.04$ & $-0.27 \ \pm \ 0.05$ & $-0.26 \ \pm \ 0.05$ & 
$-0.22 \ \pm \ 0.06$ & / \\
\bottomrule                                            
\end{tabular}
\tablefoot{Under the assumption of pure self-similarity, the scaling residuals $\delta\alpha_M$ and $\delta\alpha_z$ are expected to be identically zero. To obtain the total scaling values, the tabulated values must be added to the self-similar baseline: $\alpha_z = 2, 2/3, 8/3, -2/3$ and $\alpha_M = 0, 2/3, 2/3, 2/3$ for density, temperature, pressure, and entropy, respectively (see Eqs.~\ref{eqn:rho500}, \ref{eqn:T_P_K_500}).}
\end{table*}

\subsection{Radial variation of mass and redshift scalings}\label{subsec:radialvarimassred}

Figure \ref{fig:alpha_M_z} shows the radial variation, relative to self-similar, of the redshift ($\delta \alpha_z$) and mass ($\delta \alpha_{M}$) scaling factors for the density, temperature, pressure, and entropy profiles, obtained using the piecewise power-law fitting procedure (see Sect.~\ref{subsec:modeling}; Eq.~\ref{eqn:QQ500}).
Plotting the deviations from self-similarity this way allows one directly to assess when a quantity has a 'stronger' or 'weaker' dependence than standard.

The density redshift scaling factor profile, $\delta \alpha_z (n_{\rm e}, R)$, indicates weaker evolution than self-similar in the central regions. Around $R\sim0.5\,R_{500c}$, the evolution becomes slightly stronger than self-similar, staying roughly flat at $\delta \alpha_z (n_{\rm e})\sim0.3$. The temperature redshift scaling profile, $\delta \alpha_z (T, R)$, indicates slightly different behaviour, being consistent with self-similar in the central regions while becoming weaker than self-similar beyond $R\sim0.5\,R_{500c}$. However in both cases, the evolution factor is consistent with self-similar at $R_{500c}$ at the $3\sigma$ level.

The mass scaling profiles exhibit somewhat different behaviour. For the density, the mass scaling factor $\delta \alpha_{M} (n_{\rm e}, R)$ exhibits a strong radial variation, being significantly stronger than self-similar in the central regions, with $\delta \alpha_M (n_{\rm e}) \sim 0.4$, and only becoming self-similar around $R_{500c}$. In contrast, there is no significant radial variation in the temperature scaling, which stays flat at $\delta \alpha_M (T, R)\sim -0.15$ over the full radial range. The mass and redshift scaling profiles of pressure and entropy are fully consistent with expectations, given the density and temperature variations discussed above.

\subsection{Universal thermodynamic profiles}\label{subsec:thermod_profiles_results}

We next described the scaled thermodynamic profiles with the universal analytical models presented in Sect.~\ref{subsec:analytmodel}. Rather than using one single average redshift and mass scaling factor as in previous work \citep[e.g.][]{pra22}, we adopted the radial redshift and mass scalings obtained from a piecewise fit, evaluated in bins of $\Delta R =0.2\, R_{500c}$, and their corresponding posteriors as priors for the redshift and mass scalings when fitting the scaled profiles. The numerical values for these scalings, relative to self-similar, can be found in Table~\ref{table:mass_scale_best}. Each thermodynamic profile fit therefore contains five (six for the density) additional radial evolution and mass scaling terms. The radial profile fitting followed the methodology outlined in Sect.~\ref{subsec:analytmodel}, with priors given in Table~\ref{table:prior_and_best_fit}. The fitting process incorporates propagation of the uncertainty on the individual cluster masses, $M_{500c}$ (and hence the scaling radius, $R_{500c}$), so that the individual masses are as a natural output of our fit. All are consistent with the initial $M_{500c}$ estimates within a $2\sigma$ limit. A detailed mass investigation lies beyond the scope of this work; they are therefore marginalised over when deriving the parameter posterior distributions of the scaled thermodynamical profiles. 

\begin{figure*}
\centering
\includegraphics[width=16cm]{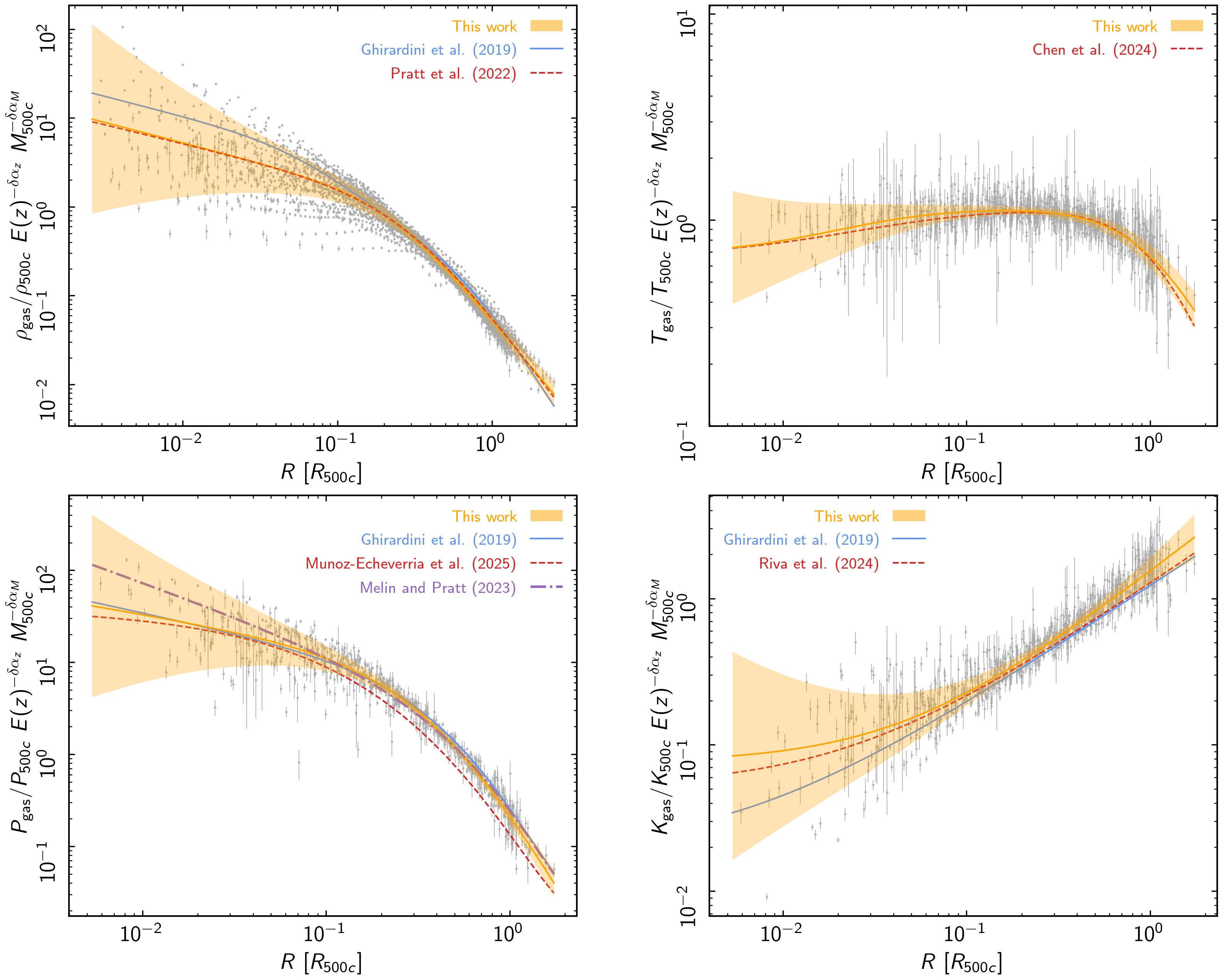} \\ 
\caption{\footnotesize Scaled ICM density (top left panel), temperature (top right panel), pressure (bottom left panel), and entropy (bottom right panel) profiles for our sample of 92 SZE-selected clusters (grey points), overplotted with the best-fitting models (solid orange line): gNFW (for density and pressure), Vikhlinin (for temperature), and power law with core (for entropy). The models include a radially varying intrinsic scatter term, shown as orange envelopes. The solid light blue lines in all panels show the results obtained by \cite{ghirardini_universal_2019} for the X-COP sample.
The dashed coloured lines indicate the the best-fitting universal profiles in \cite{pra22}, \cite{melinpratt2023}, \cite{chen_evolution_2024}, \cite{munoz-echeverria_chex-mate_2025}, and \cite{riva_chex-mate_2024}.}
\label{fig:dens_temp_best_fitt_profiles}
\end{figure*}

The resulting best-fitting parameters for the analytical model fits are given in Table~\ref{table:prior_and_best_fit}, together with the adopted priors, while Figs.~\ref{fig:cornerplot_density} and~\ref{fig:cornerplot_temperature} show the marginal posterior distributions of the best-fitting model parameters for the density and temperature profiles. 
Figure~\ref{fig:dens_temp_best_fitt_profiles} shows the scaled density, temperature, pressure, and entropy profiles of the sample, along with the best-fitting analytical models and comparisons with published results, further discussed in Sec.~\ref{subsubsec:comparison_with_literature}.

\begin{figure}
\centering
\includegraphics[width=8cm]{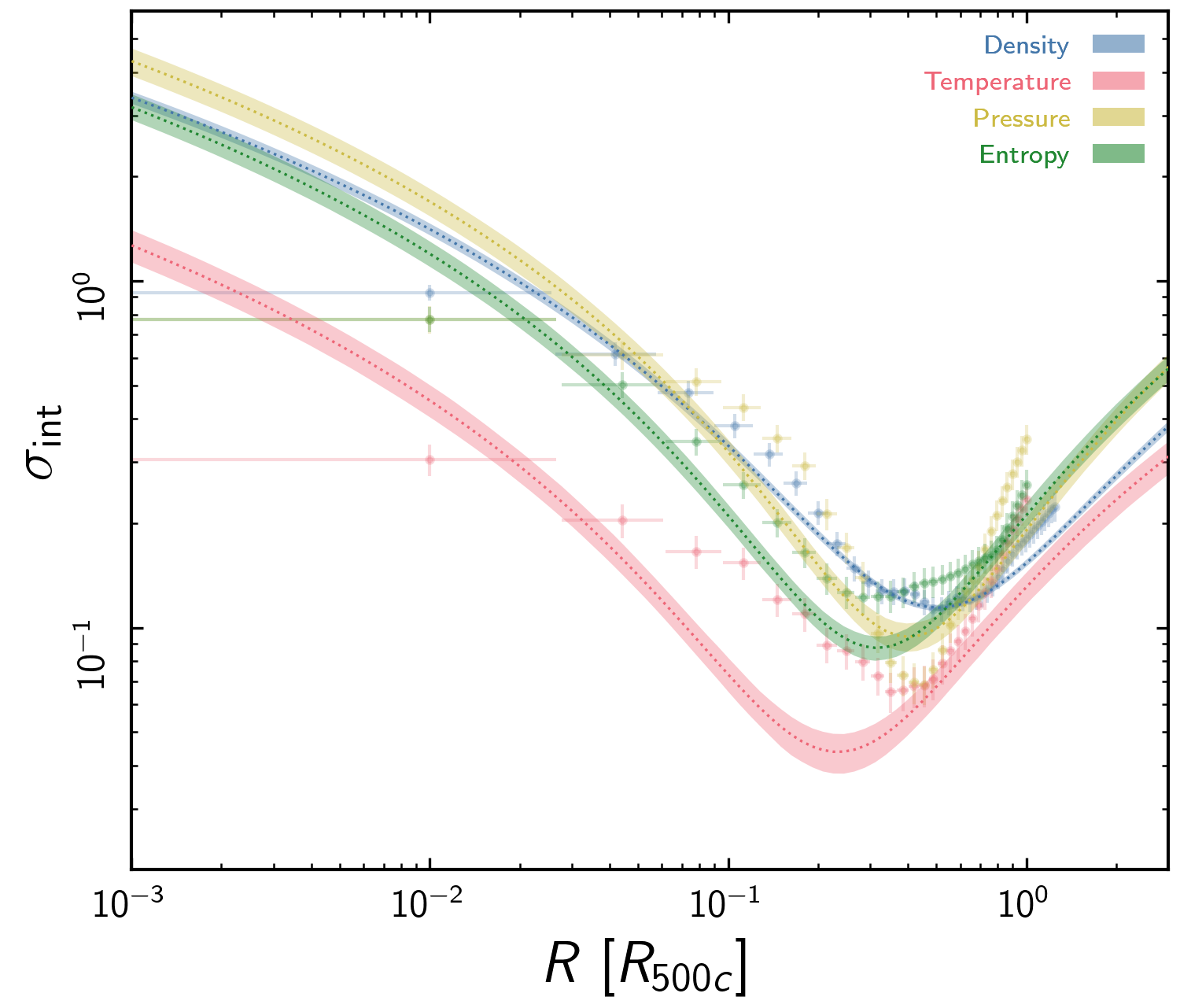} \\
\caption{\footnotesize Measured intrinsic scatter of all thermodynamic quantities considered: density (blue), temperature (red), pressure (yellow), and entropy (green) displayed for the piecewise fit  (points with error bars -- Eq.~\ref{eqn:QQ500}) and natural log-parabola analytical fit approach (dashed lines and colour envelopes -- Eq.~\ref{eqn:medianprofile}). The intrinsic scatter is defined in relative terms: a value of $0.1$ on the y-axis indicates that the considered quantity is intrinsically scattered by $10\%$ around its mean value.}
\label{fig:int_scatter_prof}
\end{figure}

Figure \ref{fig:int_scatter_prof} shows the intrinsic scatter profiles described either by the analytical parabolic form (see Eq. \ref{eqn:intrscatter}) or by the piecewise approach (see Section \ref{subsec:modeling}) for the four thermodynamic quantities considered. There is good agreement between the two approaches adopted to characterise the scatter in the density profiles. Both methods reconstruct the typical shape of the intrinsic scatter and recover the location of its minimum at $\sim 0.6 \ R_{500c}$, with a value of $15-20\%$ for all quantities. However, the functional-form description tends to overestimate the scatter in the inner regions compared to the piecewise approach, which is free of parametric assumptions. The temperature exhibits the least intrinsic scatter among the thermodynamic quantities, yet shows the largest discrepancy between the two characterisations. This is partly due to the measurement uncertainties becoming comparable to the intrinsic scatter in the temperature profiles (see Fig.~\ref{fig:dens_temp_best_fitt_profiles}).
The intrinsic scatter profiles of the pressure and entropy follow the combined trend observed for the density and the temperature. 

\begin{table}     
\caption{\footnotesize Best-fitting parameters of the functional forms describing the universal thermodynamical profiles introduced in Section \ref{subsec:analytmodel}, along with the priors used in the MCMC fitting process.}
\centering 
\textbf{Density}: Eq. \ref{eqn:GNFW} \\[3pt]
\begin{tabular}{ccc}        
\toprule\toprule
Parameter & Priors & Best fit \\
\hline
ln($n_0$) & $\mathcal{U}(-8.4,11.6)$ & $0.25^{+0.14}_{-0.13}$\\
$x_s$ & $\mathcal{U}(0.,2.5)$ & $0.22^{+0.01}_{-0.01}$\\
$\alpha$ & $\mathcal{U}(-0.01,10)$ & $0.45^{+0.06}_{-0.07}$\\
$\beta$ & $\mathcal{U}(0.01,10)$ & $0.70^{+0.02}_{-0.02}$\\
$\gamma$ & $\mathcal{U}(0.01,10)$ & $1.89^{+0.22}_{-0.20}$\\
$\textup{ln}(M_{500c,i})$ & $\mathcal{N}(\textup{ln}(M^{\rm cat}_{500c,i}), \sigma_{M_{500c}, i}^2)$ & $\sim M^{\rm cat}_{500c,i}$ \\
$\sigma_0$ & $\mathcal{U}(0,0.5)$ & $0.12^{+0.01}_{-0.01}$\\
$\sigma_1$ & $\mathcal{U}(0,0.5)$ & $0.08^{+0.01}_{-0.01}$ \\
ln$(x_0)$ & $\mathcal{U}(-5.9,0.9)$ & $-0.68^{+0.02}_{-0.02}$\\
\hline                                            
\end{tabular} \\ 
\vspace{1em}
\textbf{Temperature} Eq. \ref{eqn:tempmodl} \\[3pt]
\begin{tabular}{ccc}        
\hline\hline 
Parameter & Priors & Best fit \\
\hline
ln($T_0$) & $\mathcal{U}(-10,10.3)$ & $0.16^{+0.05}_{-0.02}$\\
$T_{\rm min}/T_0$ & $\mathcal{U}(0,1.5)$ & $0.58^{+0.18}_{-0.35}$\\
ln($x_{\rm cool}$) & $\mathcal{U}(-7,0)$ & $-3.75^{+0.65}_{-1.09}$\\
$a_{\rm cool}$ & $\mathcal{U}(0,3)$ & $1.31^{+0.90}_{-0.58}$\\
$x_{\rm t}$ & $\mathcal{U}(0.01,3)$ & $1.11^{+0.25}_{-0.26}$\\
$c$ & $\mathcal{U}(0,3)$ & $1.88^{+0.67}_{-0.56}$\\
$\textup{ln}(M_{500c,i})$ & $\mathcal{N}(\textup{ln}(M^{\rm cat}_{500c,i}), \sigma_{M_{500c}, i}^2)$ & $\sim M^{\rm cat}_{500c,i}$\\
$\sigma_0$ & $\mathcal{U}(0,0.5)$ & $0.04^{+0.01}_{-0.01}$\\
$\sigma_1$ & $\mathcal{U}(0,0.5)$ & $0.04^{+0.01}_{-0.01}$\\
ln$(x_0)$ & $\mathcal{U}(-5.1,0.6)$ & $-1.45^{+0.07}_{-0.06}$\\
\hline                                            
\end{tabular} \\
\vspace{1em}
\textbf{Pressure}: Eq. \ref{eqn:GNFWpress} \\[3pt]
\begin{tabular}{ccc}        
\hline\hline 
Parameter & Priors & Best fit \\
\hline
ln($P_0$) & $\mathcal{U}(-7.4,12.6)$ & $2.24^{+0.51}_{-0.35}$\\
$c_{500}$ & $\mathcal{U}(0,5)$ & $2.39^{+0.57}_{-0.75}$\\
$\alpha$ & $\mathcal{U}(0.01,5)$ & $0.34^{+0.15}_{-0.18}$\\
$\beta$ & $\mathcal{U}(1,10)$ & $3.60^{+0.92}_{-0.51}$\\
$\gamma$ & $\mathcal{U}(0.01,5)$ & $1.34^{+0.39}_{-0.30}$\\
$\textup{ln}(M_{500c,i})$ & $\mathcal{N}(\textup{ln}(M^{\rm cat}_{500c,i}), \sigma_{M_{500c}, i}^2)$ & $\sim M^{\rm cat}_{500c,i}$ \\
$\sigma_0$ & $\mathcal{U}(0,0.5)$ & $0.09^{+0.01}_{-0.01}$\\
$\sigma_1$ & $\mathcal{U}(0,0.5)$ & $0.12^{+0.01}_{-0.01}$\\
ln$(x_0)$ & $\mathcal{U}(-5.1,0.6)$ & $-0.91^{+0.05}_{-0.04}$\\
\hline                                            
\end{tabular} \\
\vspace{1em}
\textbf{Entropy}: Eq. \ref{eqn:powerlaw_entr} \\[3pt]
\begin{tabular}{ccc}        
\hline\hline 
Parameter & Priors & Best fit \\
\hline
ln($K_0$) & $\mathcal{U}(-7,3)$ & $-2.58^{+0.13}_{-0.15}$\\
$K_1$ & $\mathcal{U}(0.5,2)$ & $1.47^{+0.04}_{-0.03}$\\
$\alpha$ & $\mathcal{U}(0,3)$ & $0.99^{+0.04}_{-0.03}$\\
$\textup{ln}(M_{500c,i})$ & $\mathcal{N}(\textup{ln}(M^{\rm cat}_{500c,i}), \sigma_{M_{500c}, i}^2)$ & $\sim M^{\rm cat}_{500c,i}$ \\
$\sigma_0$ & $\mathcal{U}(0,0.5)$ & $0.09^{+0.01}_{-0.01}$ \\
$\sigma_1$ & $\mathcal{U}(0,0.5)$ & $0.09^{+0.01}_{-0.01}$\\
ln$(x_0)$ & $\mathcal{U}(-5.1,0.6)$ & $-1.15^{+0.05}_{-0.05}$\\
\hline                                            
\end{tabular}
\tablefoot{Uniform priors between $a$ and $b$ are indicated with $\mathcal{U}(a,b)$, and Gaussian priors centred in $\mu$ with variance $\sigma^2$ are indicated with $\mathcal{N}(\mu, \sigma^2)$. For each cluster $i$, we assume a natural log-normal distribution for the mass, based on $ M^{\rm cat}_{500c,i}$ values reported in the catalogues.}
\label{table:prior_and_best_fit}
\end{table}

\section{Discussion}\label{sec:discussion}

\subsection{Comparison to previous work}\label{subsubsec:comparison_with_literature}

We compared our results for the mass and redshift scaling profiles with those presented in \citet{mcdonald_remarkable_2017} and \citet[]{riva_chex-mate_2024}. The sample analysed in \cite{mcdonald_remarkable_2017} combines a low-$z$ X-ray selected sample \citep[27 systems at $0<z<0.1$;][]{vikhlinin_chandra_2009-1, vikhlinin_chandra_2009} with SPT-selected clusters at intermediate \citep[90 systems at $0.25<z<1.2$;][]{mcdonald_growth_2013, mcdonald_redshift_2014, nurgaliev_testing_2017} and high $z$ \citep[8 systems at $1.2<z<1.9$;][]{bleem_galaxy_2015} redshifts. \citet{riva_chex-mate_2024} analysed a sample of 32 high-mass ($M_{500c} > 7.75 \times 10^{14}\, \msol$) \textit{Planck} SZE-selected galaxy clusters at $z>0.2$, drawn from the CHEX-MATE sample \citep[][]{collaboration_cluster_2021, rossetti_chex-mate_2024}. In addition to those clusters, they considered the \rexcess \ X-ray selected sample (see Sect.~\ref{subsec:SZDataset}), and the {\it Planck} ESZ sample \citep[][]{collaboration_planck_2011}, which comprises 62 SZE-selected clusters at $z<0.5$, spanning one decade in mass. 

\begin{figure}
\centering
\includegraphics[width=8cm]{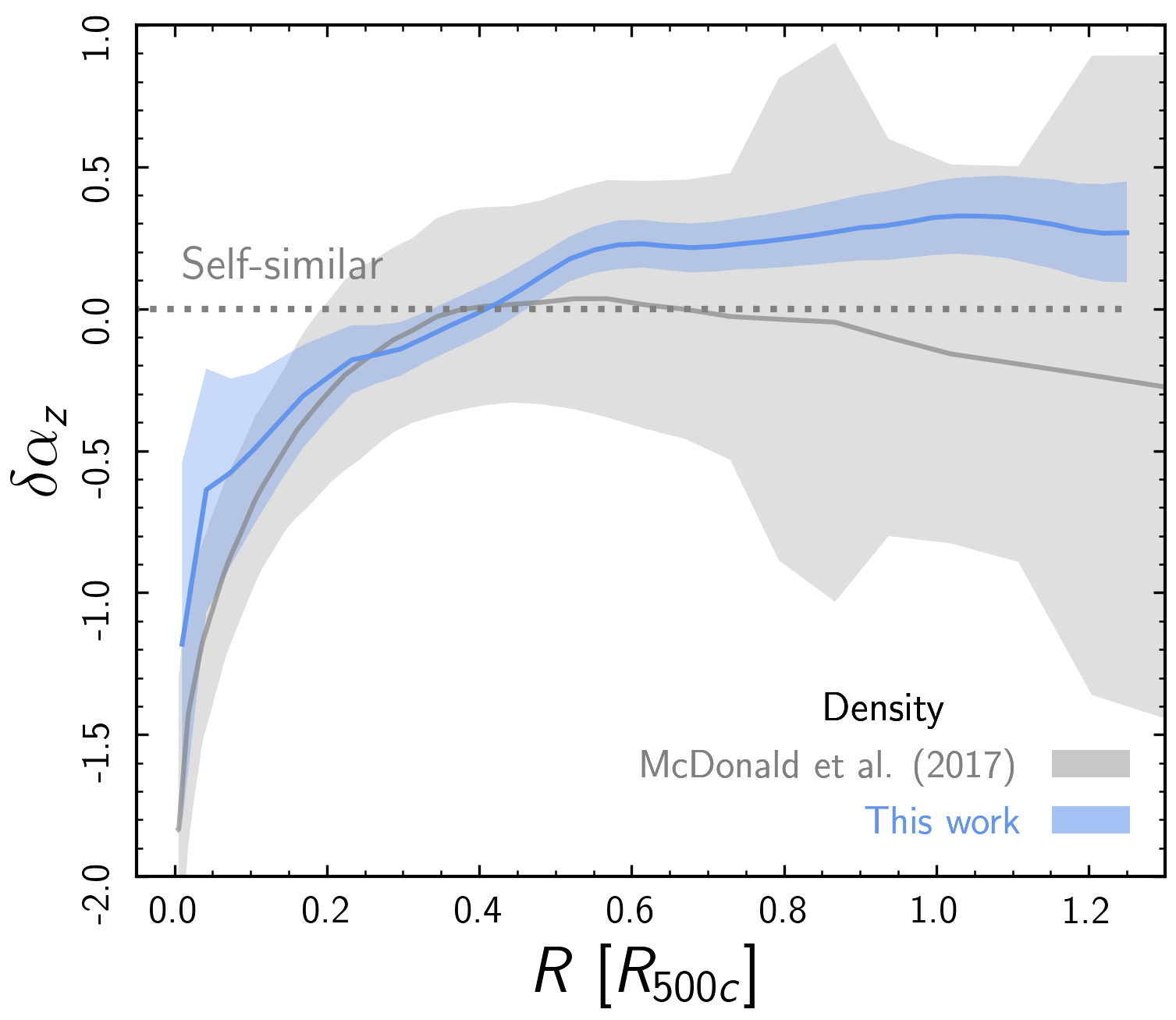} \\ 
\caption{\footnotesize Profiles for the evolution scaling factor for the density profile, shown for the sample studied in \cite{mcdonald_remarkable_2017} (in grey) and ours  (blue).
The shaded areas indicate the corresponding $1\sigma$ uncertainties. }
\label{fig:Mcdonald_Riva_compar}
\end{figure}

Our analysis provides much tighter constraints on the radial dependence of the evolution scaling factor for the density, $\delta\alpha_z (n_{\rm e}, R)$, compared to the results reported in \citet[][see Fig.~\ref{fig:Mcdonald_Riva_compar}]{mcdonald_remarkable_2017}.

\begin{figure}
\centering
\includegraphics[width=8cm]{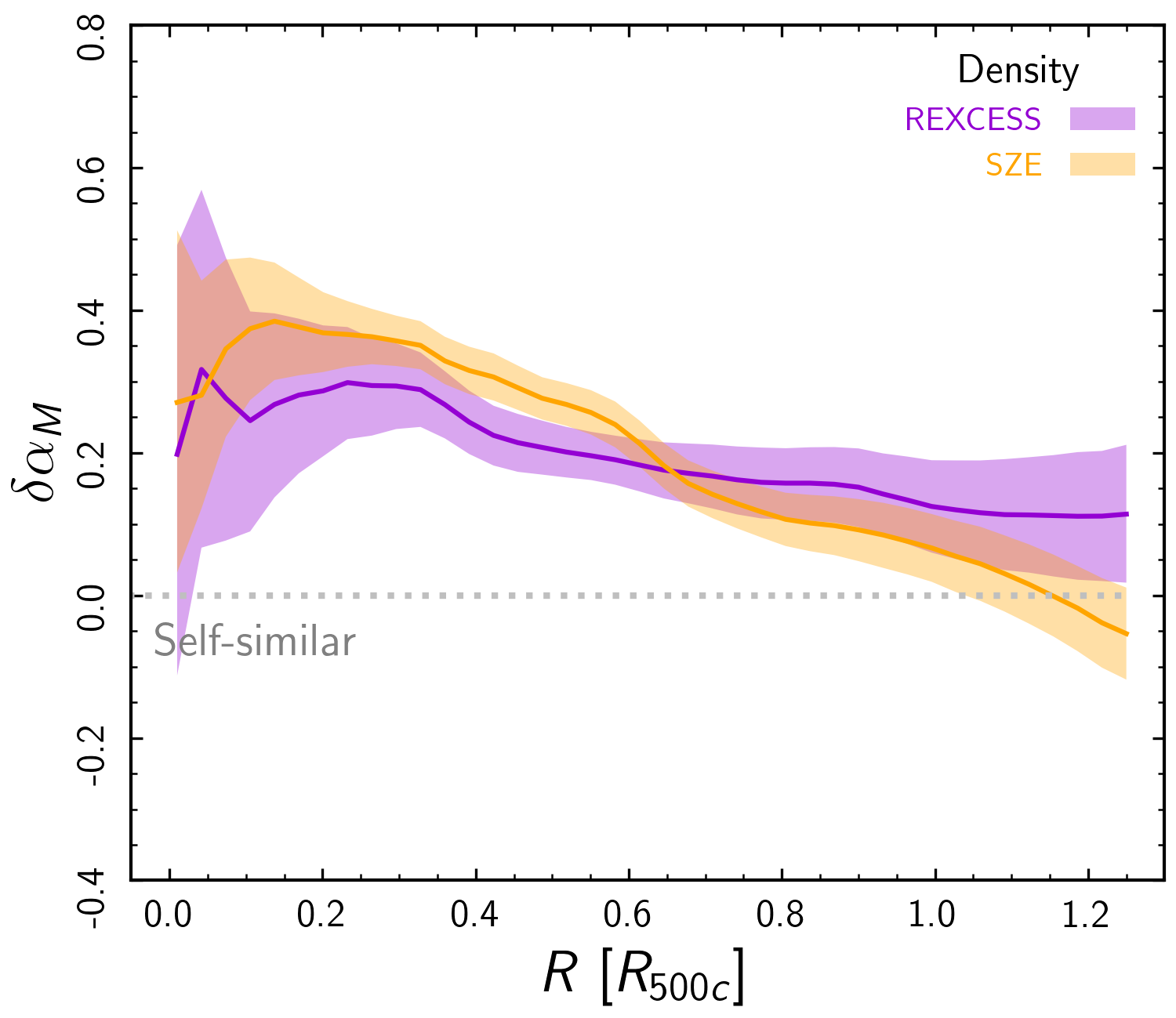} \\ 
\includegraphics[width=8cm]{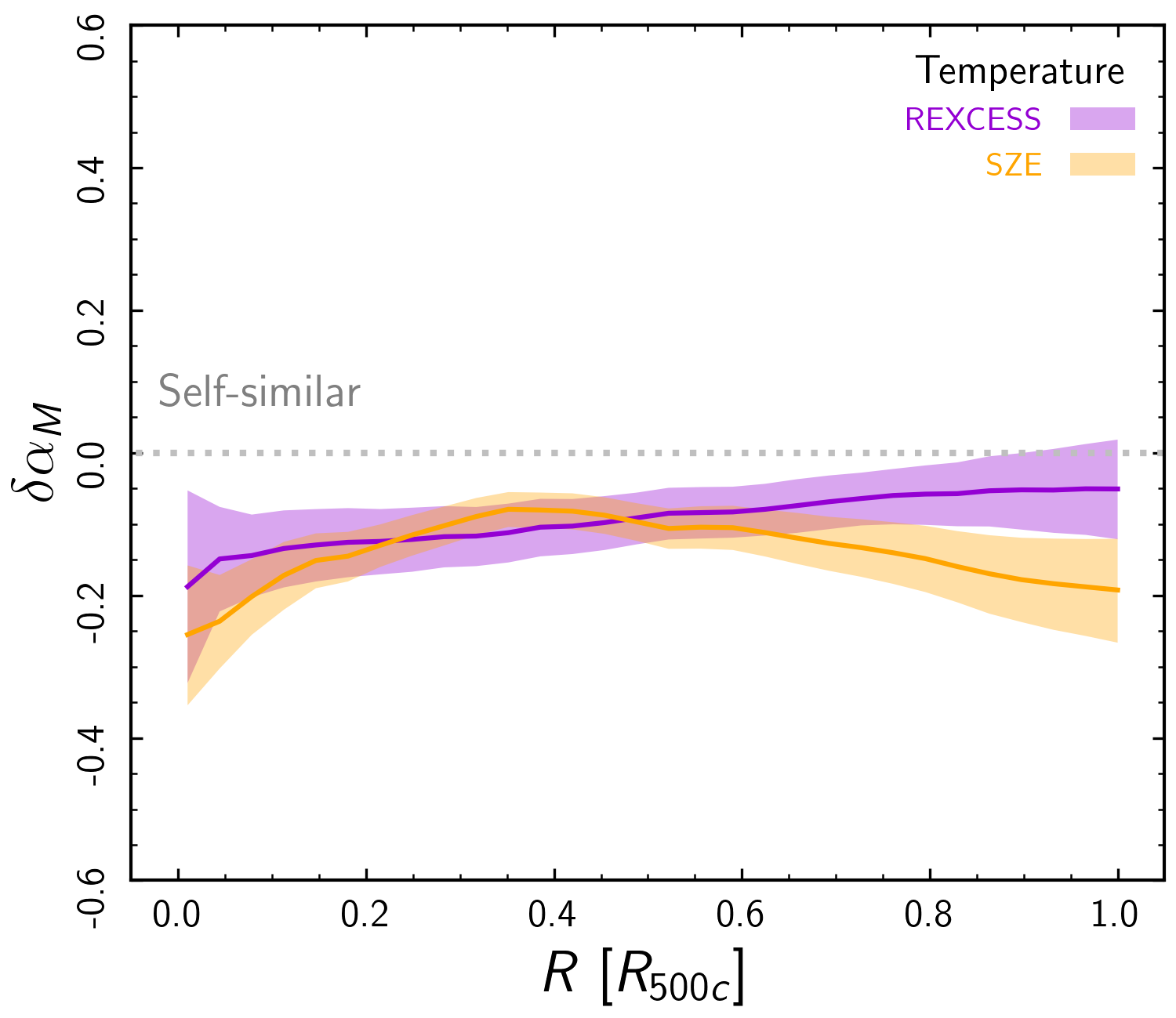} \\ 
\caption{\footnotesize Radial profiles of the mass scaling for density (top panel), temperature (bottom panel) profiles for our 92 SZE-selected clusters (orange) and the \rexcess\ sample (purple), with their $1\sigma$ uncertainty envelopes.} 
\label{fig:alpha_M_Xray_SZ}
\end{figure}

To investigate differences in mass trends between X-ray and our SZE-selected samples, we undertook a comparison of the mass and redshift scaling profiles obtained from the 92 SZE-selected clusters with those derived from the X-ray selected \rexcess \ sample. For the latter, we applied the same fitting procedure described in Sect.~\ref{subsec:modeling}, but fixed the evolution scaling factor to its predicted self-similar value, as the limited redshift range of the \rexcess\ sample does not allow for a meaningful constraint to be put on this parameter. Figure \ref{fig:alpha_M_Xray_SZ} shows the resulting deviations relative to self-similar of the mass scaling factors of the density, $\delta \alpha_M (n_{\rm e}, R)$, and temperature, $\delta \alpha_M (T,R)$, for the \rexcess\ density and temperature compared to those for our SZE-selected sample. The minor deviations observed remain within the associated uncertainties. Given that \rexcess\ constitutes a representative sample of the low-redshift cluster population, these results demonstrate that the observed mass trends are robust and independent of the sample selection method.

We then compared the best-fitting results for the scaled thermodynamic profiles with recent work that has accounted for an intrinsic scatter term (see Figure \ref{fig:dens_temp_best_fitt_profiles}). \citet{ghirardini_universal_2019} analysed the thermodynamic properties of the X-COP sample, comprising the most massive nearby clusters ($0.04 < z < 0.1$ and $\rm M_{500c}>5 \times 10^{14}\, M_{\odot}$) detected in the \textit{Planck} survey. 
Using an approach similar to ours, they derived universal profile shapes (see Section \ref{subsec:analytmodel}), using a quadratic functional form for the radial intrinsic scatter term (see Eq. \ref{eqn:intrscatter}). However, they did not account for any deviations in mass and redshift from self-similarity. Despite the different statistical significance of the individual samples, there is good consistency between our results and those from X-COP \citep[see Table 3 in][]{ghirardini_universal_2019}. Their qualitative description of the density profiles with a \cite{vikhlinin_chandra_2006} functional form is consistent with our characterisation at all radii (see Figure \ref{fig:dens_temp_best_fitt_profiles}). We omit a direct comparison of our 3D deprojected temperature results, as the temperature profiles provided by \citet{ghirardini_universal_2019} were projected (2D). Both pressure and entropy are also in good agreement.

Using joint observations from \planck\ and SPT \cite{melinpratt2023}, constrained SZ radial pressure profiles of 461 galaxy clusters. Their best-fitting universal pressure profile with a gNFW model is consistent with ours at $R>0.1 \ R_{500c}$; the differences at smaller radii are likely due to the lower angular resolution of SZ observations compared to our X-ray data.

\cite{munoz-echeverria_chex-mate_2025} recently performed a similar analysis on the combined scaled X-ray (\xmm) and SZ (\planck) pressure profiles for a subsample of 28 galaxy clusters from the CHEX-MATE sample \citep[DR1,][]{collaboration_cluster_2021, rossetti_chex-mate_2024}, spanning mass and redshift ranges of $[2-14] \times 10^{14}\, M_{\odot}$ and $0.07 < z < 0.6$, respectively. 
They obtained a value of $\delta\alpha_M (P)$ consistent with zero from the marginalised posterior distribution, while their best-fitting model yielded $\sim 0.2$. The range spanned by these two estimates is broadly consistent with our radially averaged estimate of $\langle \delta\alpha_M (P) \rangle \simeq 0.12$, as well as with the results of our global power-law fit, in which the radial pressure profile is described by a gNFW model (Eq.~\ref{eqn:GNFWpress}) over the radial range $[0,1]\ R_{500c}$, assuming a single mass and redshift scaling. This fit yields $\delta\alpha_M (P) = 0.12 \pm 0.02$ and $\delta\alpha_z (P) = -0.07 \pm 0.05$.
This reinforces the conclusion that the observed mass deviations, significant at the $3\sigma$ level over $\sim 0.2-0.6 \ R_{500c}$, are a consistent property across independent galaxy cluster samples.  Their pressure profiles leverage both X-ray and SZE observations, providing coverage across a wide radial range; despite this, our best-fitting parameters are in good agreement (within $1\sigma$) with those reported in their study. The notable difference appears in the outer shape of the intrinsic scatter profiles (bottom left panel of Fig.~ \ref{fig:dens_temp_best_fitt_profiles}). This might be associated with differences in the functional form adopted to describe the intrinsic scatter, the radial coverage -- as \planck\ measurements extend out to $\sim 2.5$~$R_{500c}$ -- or the increasing contribution of statistical uncertainties to the total error budget for the SZ data at large radii. Lastly, we note that our best-fitting mean, $\delta\alpha_M (P)$, is also in agreement with that found by \citet{arnaud_universal_2010} for the X-ray-selected \rexcess\ sample.

\cite{riva_chex-mate_2024} described the entropy profiles of a CHEX-MATE sub-sample (see Sect.~\ref{subsubsec:comparison_with_literature}) using the same universal and intrinsic scatter models presented in Eqs.~\ref{eqn:powerlaw_entr} and~\ref{eqn:intrscatter}. Although their best-fitting parameters and corresponding statistical uncertainties do not fully agree with those obtained in this work, the overall intrinsic scatter profile remains consistent with ours, as shown in the bottom right panel of Fig.~\ref{fig:dens_temp_best_fitt_profiles}. 

At higher redshifts, a comparison of our density profiles with those from \cite{pra22} shows excellent agreement. Similarly, our temperature results are fully consistent with the findings of \cite{chen_evolution_2024} for the X-ray selected sample \excpres, which covers a mass and redshift range of $[1-13] \times 10^{14} \, M_{\odot}$ at $0.4<z<0.6$. This suggests that while a more precise characterisation of the mass and redshift scalings reduces intrinsic scatter in the scaled profiles, it does not fundamentally change the average scaled profiles. 

Finally, we note that predictions evaluated at $R_{500c}$ based on the semi-analytical model proposed by \cite{ettori_icmz_2023} give $\delta\alpha_M = 0.23, -0.23, -0.16 \ \pm 0.01$ and $\delta\alpha_z = 0.11, 0.11, -0.07 \ \pm 0.03$, for density, pressure, and entropy profiles, respectively. This model does not account for any radial variation in the departure from self-similarity, explaining the discrepancies with our results.  
However, their predictions are consistent with our radial averages of $\delta\alpha_M$ and $\delta\alpha_z$ profiles, showing that the global scaling aligns.

\subsection{Impact of the mass scale proxy and clumping factor}\label{subsubsec:mass_def_dependence} 

Our baseline analysis uses mass estimates derived from the $Y_X-M_{500c}$ relation \citep[$M_{Y_X}$;][]{kravtsov_new_2006}. To quantify how the mass proxy choice affects our mass scaling profiles, we defined a general linear relation between an alternative estimate, $M_{A}$, and our reference: $\ln{M_{Y_X}} = m \cdot \ln{M_{A}} \ + \ c \ \pm \sigma$. We then introduced three perturbations: 
\begin{itemize}
\item a constant $20\%$ downward bias ($c=-\textup{ln}\,0.8$, $m=1$), equivalent to a variation in the normalisation of the absolute mass scale, chosen to mimic a constant hydrostatic bias;
\item a mass-dependent slope ($c=\textup{ln}\,M_p^{0.2}$, $m = 0.8$) anchored at $M_p=5\times10^{14} M_{\odot}$, equivalent to a mass-dependent variation in the relation between the mass proxy and the reference mass;
\item an intrinsic scatter of $20\%$ ($m=1$, $\sigma = \textup{ln}\,1.2$), corresponding to a large scatter between the mass proxy and the true mass;
\end{itemize}
and quantified their effects on the resulting redshift and mass scaling profiles. Beyond these individual effects, we also evaluated a general case incorporating all three perturbations simultaneously. For each case, we applied the model described in Sect.~\ref{subsec:modeling}, allowing all four parameters to vary freely. 

\begin{figure}
\centering
\includegraphics[width=8cm]{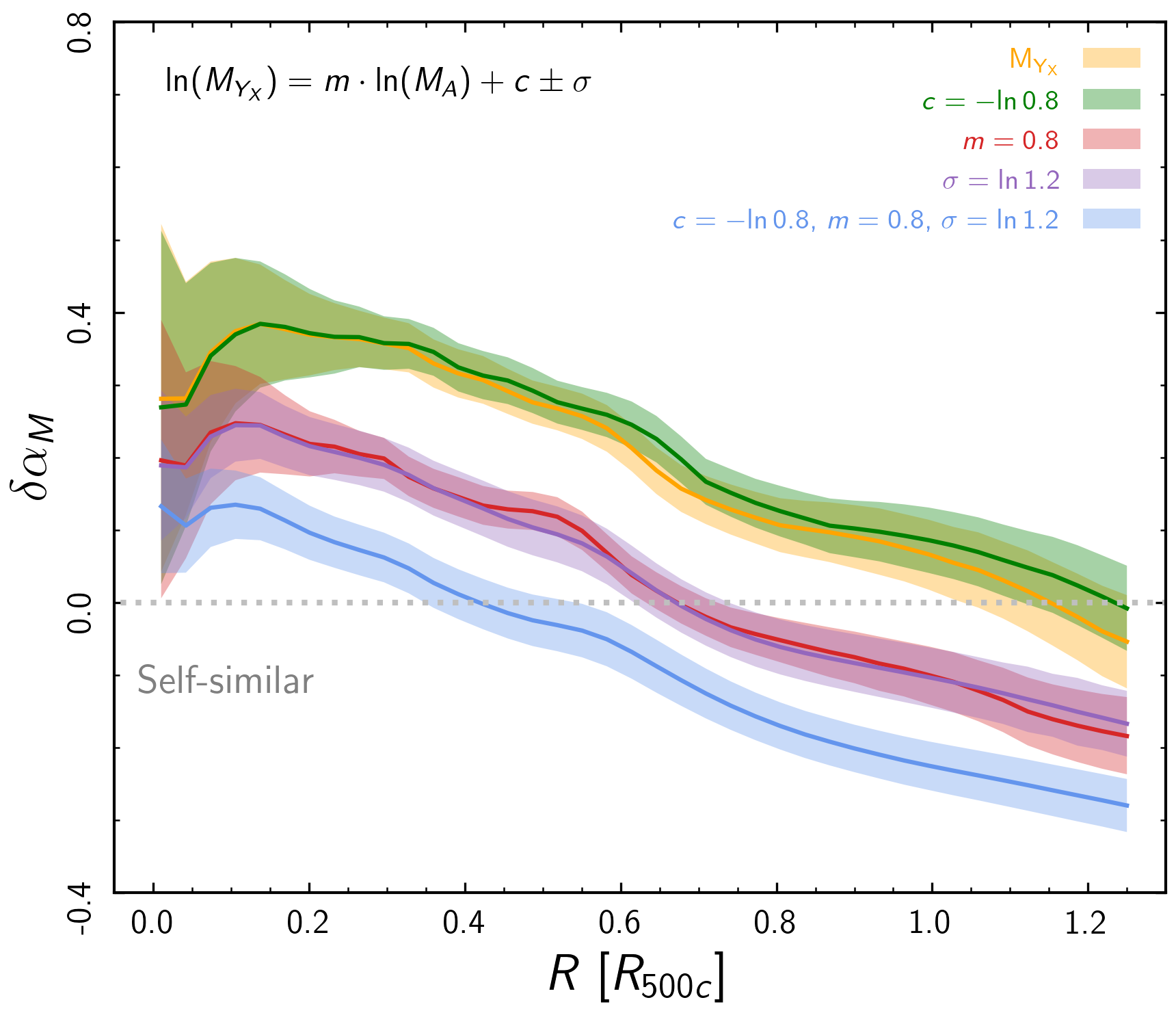} \\ 
\caption{\footnotesize Mass scaling factor profiles derived from scaled density profiles for various mass-dependence configurations: $M_{500c,Y_X}$ reference (orange); constant downward bias of $20\%$ in mass (green); mass-dependent bias with a slope of $m=0.8$, fixing the value of the pivot mass (red); intrinsic scatter of $20\%$ (purple). The light blue curve shows the cumulative effect of the three perturbations.}
\label{fig:mass_estimates_test}
\end{figure}

No significant differences are observed in the redshift scaling profiles; as this was expected, they are omitted from this work. However, Fig.~~\ref{fig:mass_estimates_test} shows that the resulting mass scaling factor profiles are not equally sensitive to these variations. While a constant offset between the mass proxy and the true mass has no effect on the profiles, both a mass-dependent bias and a larger intrinsic scatter primarily shift the normalisation, again without distorting the profile shape.
This effect on the amplitude of the radial deviations is observed for all thermodynamic quantities: we display the analysis for the density profiles in Fig.~\ref{fig:mass_estimates_test}.

The stability of the mass scaling factor profile shape to these effects suggests that the radial mass dependence we find is robust and constitutes an intrinsic feature of the cluster population. However, while the shape of the radial deviations from self-similarity are independent of the absolute mass scale, their amplitude may be affected by a strong mass dependence in the relation between the mass proxy and the true mass, or by increased scatter in this relation. Such effects could conceivably arise from the differential impact of non-gravitational processes at low and high masses, increased scatter due to the influence of dynamical state, or a variation in the dynamical state with mass. These effects would be interesting to investigate further with numerical simulations. 

We note that recent results from \cite{andreon2026} show a weaker mean mass dependence for pressure profiles when rescaled by weak-lensing masses ($\delta \alpha_M(P) \simeq -0.11$). From the above investigation, this may be due to the large scatter known to be present in the weak lensing mass with respect to the true mass, or a mass dependence of the relation between the weak lensing mass and true mass. This result may also have a bearing on the comparison with \citep{munoz-echeverria_chex-mate_2025} discussed above, as they used Gaussian priors centred on dynamical mass estimates \citep{sereno25}, which also exhibit a large scatter with respect to the true mass, as the basis of their mass scale.

Finally, unresolved clumps in our density profiles can lead to an overestimation of the measured local densities \citep{nagailau2011, zhuravleva2014}, potentially biasing the observed thermodynamic quantities. The standard clump correction adopted in the literature reconstructs the radial density profiles from the median surface brightness distribution measured in concentric annuli around the cluster centre, applying a Voronoi tasselation algorithm \citep{cappellari2003, diehlstatler2006}. Given the high level of correlation between cells that this algorithm would produce for most high-redshift clusters, we instead adopted the density correction factor of $-7\pm5 \%$ at $R_{500}$ measured in a sample of ROSAT clusters in \cite{eckert_gas_2015}, and a pressure bias of $\pm 3\%$ at $R_{500c}$ from simulations by \cite{khedekar2013}. These translate into a $7\pm6 \%$ bias on temperature, as predicted theoretically by \cite{avestruz2016}, and a $12\pm 9 \%$ bias on entropy. Propagating these biases through 100 Gaussian realisations of the profiles and repeating the fitting procedure described in Section \ref{subsec:modeling} at $R_{500c}$, we find the resulting systematic uncertainties on $\delta\alpha_z$ and $\delta\alpha_M$ to be $\sim4-10$ times smaller than the corresponding statistical uncertainties. 
For the density profiles, however, adopting the clumping factor measured by \cite{eckert_gas_2015} in the outermost radial bin, $[1,1.25] \ R_{500c}$, would yield systematic uncertainties comparable to the statistical ones, increasing the total uncertainties by approximately $20\%$ when added in quadrature. This would not affect our main conclusions, and we therefore consider clumping-related systematics negligible in the following.

\subsection{Intrinsic scatter}\label{sub:intrinsic scatter}

\begin{figure*}
\centering
\includegraphics[width=14cm]{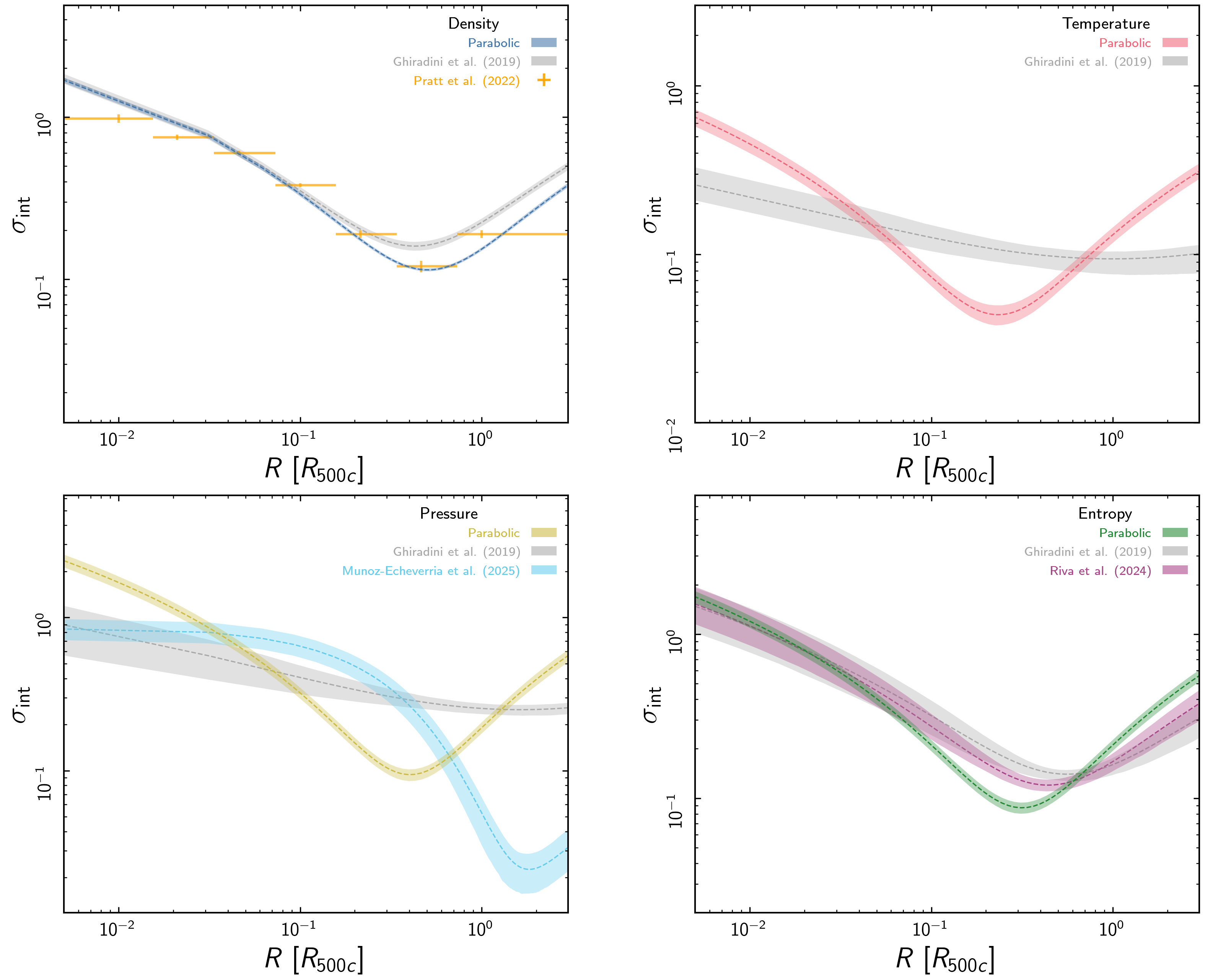} \\
\caption{\footnotesize Comparison of our derived  intrinsic scatter profiles of all thermodynamic quantities considered, density (top left panel), temperature (top right panel), pressure (bottom left panel), and entropy (bottom right panel) with previously published results. The dotted lines represent the best-fitting profiles, while the shaded areas indicate the $1\sigma$ uncertainties.
}
\label{fig:comparison_int_scatt}
\end{figure*}

In Figure~\ref{fig:comparison_int_scatt} we compare our  intrinsic scatter measurements with those in the literature. For each thermodynamic quantity considered, we find a systematically lower minimum intrinsic scatter in our sample compared to those reported in \citet{ghirardini_universal_2019}. These differences are more pronounced in the temperature and pressure profiles than in those for density and entropy. Such discrepancies are expected, as our analysis aims to disentangle the intrinsic scatter from possible additional mass and redshift scaling factors. While \citet{ghirardini_universal_2019} report lower scatter values for temperature and pressure in cluster cores, this likely stems from the shape of the adopted functional form and the specific mass and redshift range of their sample. 
With the exception of the innermost bins, our parabolic scatter description of density profiles is fully compatible with the non-analytical form of \cite{pra22}, which is defined across seven log-spaced points. In contrast, the exponential functional form adopted by \cite{munoz-echeverria_chex-mate_2025} for the pressure scatter fails to accurately describe our data. For entropy, our results show no significant discrepancy with those of \cite{ghirardini_universal_2019} and \cite{riva_chex-mate_2024}, except for a slightly deeper minimum of $\sim10\%$ reached at $R\sim 0.3 \ R_{500c}$. 

These comparisons demonstrate that both the parabolic functional form -- introduced in \cite{ghirardini_universal_2019} and adopted in this work and in \cite{riva_chex-mate_2024} -- and the exponential form from \cite{munoz-echeverria_chex-mate_2025} lack the flexibility to simultaneously reproduce the intrinsic scatter in both the inner regions and the outer radii. The radial piecewise approach, being free of parametric assumptions, is therefore expected to provide a more robust characterisation of the intrinsic scatter than any fixed functional form. While more flexible functional forms could better represent the piecewise reconstruction, exploring such alternatives is left for future work. 

The large ($>30\%$) intrinsic scatter observed in the core regions is consistent with mechanical AGN feedback and gas cooling \citep{gaspari_cause_2012, gaspari_shaken_2018}, which drive significant thermodynamic perturbations in the innermost cluster regions. Towards the outer radii, the increasing scatter is likely driven by mass accretion \citep{vazza2011, nelson2014}. The dynamical state of the clusters further contributes to the scatter in both regions, as merging or disturbed systems are known to exhibit systematically larger profile variations \citep{lovisari2017}. Finally, in the outer radii, the scatter is additionally affected by uncertainties in the background subtraction, which become increasingly significant at large distances from the galaxy cluster centres \citep[e.g.][]{reiprich2013}. Finally, we note that the covariance between $M_{\rm gas}$ and $R_{500c}$ \citep{pra22} is not accounted for in this work, which is instead focused on comparisons with previous results that similarly do not account for this effect.

\subsection{Implications for scaling relations}\label{sub:implicat_scaling_relations}

An alternative way of expressing the above results is as a set of scaling relations that have been evaluated in various radial shells.  We assessed the impact of the derived mass and redshift self-similar deviations on the $L_{\rm bol}-T$, $L_{\rm bol}-M$, $M-T$ scaling relations, which are widely employed to provide mass estimates for large cluster samples out to high redshift \citep[e.g.][]{maughan_deep_2007}. 
The X-ray luminosities were measured in the bolometric ($[0.01-100]$ keV) band from the $[0.3-2]$ keV surface brightness profile count rates and the gas temperature, $T_{\rm spec}$, estimated over the radial range considered from the density and deprojected temperature profiles as a 'spectroscopic-like' temperature \citep[see ][]{mazzotta2004}. The luminosities were derived assuming a thermal model for a given gas density and temperature profile, with the metallicity fixed to $0.3 \ Z_{\odot}$ following \cite{and89}. Uncertainties on the luminosity values account for the errors on the spectral model, the count rates, the value of $R_{500c}$ and $T_{\rm spec}$, and were estimated from Monte Carlo realisations in which the luminosity was recomputed for 100 realisations of the surface brightness profile. 
The scaling relations in each shell were described using a double power-law model (Eq. ~\ref{eqn:QQ500}) following the fitting procedure presented in Sec. \ref{sec:fitting_procedure}. 

We first considered the integrated relations ($R < R_{500c}$). These yield results in good agreement with previous literature where the samples are statistically representative and span a comparable range in mass and redshift to our own \citep[e.g.][]{pratt_galaxy_2009, bulbul2019, ettori2020, lovisari2020, pra22}. As found in previous work, our $L_{\rm bol}-T$ and $L_{\rm bol}-M$ relations exhibit slopes that are significantly steeper than self-similar predictions, while displaying a milder redshift evolution. The $M-T$ relation yields a slope compatible with self-similar expectations, albeit with a shallower redshift evolution, consistent with previous studies \citep[e.g.][]{arn05,lovisari_scaling_2022}. In all three cases, however, significant levels of intrinsic scatter are observed.

To reduce the scatter and improve the consistency of scaling relations determined for different samples, core-excised relations ($[0.15 - 1]\,R_{500c}$) are typically preferred. Our results reinforce the established consensus that excluding the central regions of galaxy clusters yields closer agreement with self-similar predictions. In this case too, we find excellent agreement with previous literature, alongside a significantly reduced level of intrinsic scatter in the scaling relations. 

Comparing the three radial shells discussed above, we find the closest agreement with self-similar expectations in the core-excised shell ($[0.15 - 1]\,R_{500c}$). In this context, however, better agreement with self-similar predictions does not purely imply that simple gravitational physics applies; rather, it emerges from a compensation or balance, in this radial range, between baryonic processes that strongly affect the core and the inhomogeneous accretion and non-equilibrium processes prevalent towards the outer radii. In other words, the core-excised shell effectively averages out the radially differentiated imprints of these processes.

\subsection{Physical interpretation}\label{sub:physical_explana}

Our results have significant physical implications. We first investigated the potential connection between the observed scaling deviations and the dynamical state of the clusters, dividing the sample into three subsamples: cool-core clusters, morphologically disturbed clusters, and clusters classified as neither relaxed nor disturbed (see Appendix \ref{app:morphological classification}). 
No statistically significant differences were found in the mass and redshift dependencies relative to self-similar predictions across the different morphological subsamples. However, this result is not strongly constraining: the relative statistical uncertainties on $\delta\alpha_M$ exceeded $25\%$ in all subsamples. These uncertainties are larger than the upper limit on the difference in the non-thermal pressure fraction between disturbed and regular clusters predicted by simulations \citep[up to $15\%$;][]{nelson2014}.
Our data are therefore uninformative concerning the impact of dynamical state on the mass and redshift dependencies, and we cannot rule out a morphology-dependent effect of this magnitude. To measure the amplitude of this effect, we would need subsamples similar in size and in redshift distribution to the full sample of $\sim 100$ clusters at $z\lesssim 1.15$.

The radial trends observed for $\delta\alpha_z (R)$ and $\delta\alpha_M (R)$ have direct physical implications for cluster evolution. The results for $\delta\alpha_z (R)$ imply that the core and the bulk do not evolve at the same time. The core appears to form early, with its subsequent evolution regulated by AGN feedback and radiative cooling. Concurrently, the cluster bulk exhibits gas that is systematically cooler and denser than self-similar predictions. This indicates that the bulk is still assembling at later epochs, consistent with ongoing, gravitationally driven accretion.

Simulations \citep{nelson2014, lau2015} predict that when rescaling the thermodynamic profiles with respect to $R_{200m}$ (see Appendix ~\ref{app:mean_matt_dens}), the non-thermal pressure fraction reduces by a factor of two, which has a direct impact on the redshift scaling factor measured for the thermodynamic profiles. We tested this prediction (Appendix ~\ref{app:sub:applic_alpha_z_r200m}) in comparison to the scaled profiles by $R_{500c}$. Although we observe a closer agreement with self-similarity in the redshift scaling profiles at $R\sim 0.6 \ R_{200m}$  for temperature and entropy, we do not observe a measurable difference for density. We find a stronger deviation for pressure for $R> 0.5 \ R_{200m}$. However, as our measured thermodynamic profiles reach only $\sim 0.6\ R_{200m}$, we cannot fully test the theoretical predictions towards outer regions.

The results for $\delta\alpha_M (R)$ alone imply that the gas content in clusters exhibits both radial and mass dependencies. The AGN feedback more efficiently redistributes the central gas out to larger radii in lower-mass systems, owing to their shallower gravitational potential wells relative to high-mass systems, thereby reducing gas densities within their core regions. This behaviour is tightly linked to the hierarchical growth of structure predicted by the standard cosmological model \citep{mccarthy2017, ayromlou2023}.

The observed variation in gas content with mass and radius has direct implications for the integrated gas mass fraction profiles, defined as $f_{\rm{gas}}(<R) = M_{\rm{gas}}(<R)/M(<R)$. To derive this quantity, we integrated the gas density  profiles to obtain the corresponding gas mass distributions, while the total mass profiles were modelled assuming a NFW profile with a concentration parameter, $c_{500c} = 3.2$ \citep{pointecouteau2005}. The gas mass fraction was then parametrised according to Eq.~\eqref{eqn:QQ500}. For clarity, within this section, the additional mass and redshift scalings for $f_{\rm gas}$ are denoted as $\delta\beta_M$ and $\delta\beta_z$, respectively. A pronounced mass dependence is observed in the gas mass fraction, with more massive systems exhibiting higher $f_{\rm gas}$ values. The measured $\delta\beta_M$ and $\delta\beta_z$ are in excellent agreement with the corresponding parameters, $\delta\alpha_M$ and $\delta\alpha_z$, derived from the density profiles, as predicted for instance  by \cite{ettori_icmz_2023}. Furthermore, for the most massive clusters ($M_{500c}>7 \times10^{14} \rm M_{\odot}$) the scaled gas mass fraction measured within $R_{500c}$ converges to the universal baryon fraction, $\Omega_b/\Omega_m = 0.16$ \citep{planck_cosmo}, at a significance level of $3\sigma$.

These results support a physical interpretation in which a combination of gentle AGN feedback, cooling, and merger-induced mixing together drives redistribution of the intracluster gas. This manifests as net gas inflows in the cluster outer radii, consistent with, for example, the recent numerical simulations of \cite{pontzen2026}, while the central regions exhibit a quasi-steady balance between inflows (cooling) and outflows (AGN feedback). In high-mass systems the outflowing gas in the outer radii remains predominantly gravitationally bound, leading to a progressive accumulation of material at intermediate radii, where it becomes denser and cooler over time.

\section{Conclusions}\label{sec:conclusion}

We examined the departures from self-similarity of the ICM  thermodynamic profiles (density, temperature, pressure, and entropy) for a sample of 92 SZE-selected clusters spanning a redshift range of $0.08<z<1.13$ and a mass range of $M_{500c}=[0.5-20]\times10^{14} \ \rm M_{\odot}$. The sample characteristics allow us to significantly extend work that has been largely confined to high-mass, local systems. We found the following results:
\begin{itemize}
    \item Using a double power-law fit to the thermodynamic quantities ($n_{\rm e}, kT, P, K$) extracted in radial shells, we found a radially dependent deviation from self-similar expectations of the redshift and mass scaling factors. These deviations are strongest in the core ($ R \lesssim0.02 \ R_{500c}$) but remain evident in the outer radii, indicating that the thermodynamic properties of galaxy clusters evolve with both redshift and mass not only in the core regions, where baryonic effects are important, but also in the bulk, where gravitational processes dominate.
    \item Both density and temperature exhibit clear radial deviations from self-similar behaviour in mass and redshift. In contrast, their combination into pressure largely cancels these trends: outside the core, the pressure profile scaling remains relatively close to self-similar, with a minimal mass dependence and a redshift evolution fully consistent with self-similar expectations. This confirms pressure as the most self-similar thermodynamic quantity, as predicted by \cite{nagai_effects_2007} and confirmed observationally by \cite{arnaud_universal_2010}, with direct implications for its use in cosmological analyses.
    \item Our analytical framework, employing piecewise mass and redshift scalings and logarithmic parabolic scatter, provides a robust description of the scaled thermodynamic profiles. The resulting radial scaling parameters are consistent with our bin-by-bin priors within $2\sigma$, while still deviating from self-similar predictions at $3\sigma$ over the $\sim 0.2-0.6 \ R_{500c}$ range. Our universal profile fits are broadly consistent with previous literature \citep[][]{ghirardini_universal_2019, pra22, riva_chex-mate_2024, chen_evolution_2024, munoz-echeverria_chex-mate_2025}.
    \item We examined the intrinsic scatter of the scaled radial profiles using both parabolic and piecewise parametrisations, finding that the dispersion peaks in the cluster cores (where the parabolic form tends to overestimate it) and declines to a minimum of $\lesssim20\%$ for all the thermodynamic quantities at $\sim 0.2$--$0.8 \ R_{500c}$. The overall consistency of our density and entropy results with the literature underscores the robustness of our universal profiles. Although temperature and pressure exhibit larger discrepancies, the inclusion of additional radial mass and redshift scaling factors clearly enhances the decomposition and characterisation of the intrinsic scatter.
    \item We assessed the impact of adopting different mass estimates relative to our fiducial $M_{500c,Y_X}$ values. While a constant change in the absolute mass scale has no effect, a mass dependence between the mass proxy and the true mass, or increased intrinsic scatter between the mass proxy and the true mass, can significantly affect the amplitude of the radial deviations from self-similar scaling but not the radial trend. Therefore, while a constant hydrostatic bias will not change our conclusions, mass-dependent biases between mass estimates or significantly increased scatter owing to differing dynamical states will affect the amplitude of the deviations we have derived. 
    \item We compared our results with those derived from the representative X-ray-selected sample \rexcess, and the results of \cite{mcdonald_remarkable_2017} and \cite{riva_chex-mate_2024}. The agreement within $1\sigma$ suggests that the observed deviations from self-similarity are a robust physical feature, independent of sample selection methods. 
    \item The impact of dynamical state and morphological classification on the mass and redshift scaling is not robustly constrained with the current sample. The mass, redshift, and radial dependence of gas content are in good agreement with those derived from the gas density profiles, as analytically predicted by \cite{ettori_icmz_2023}. Furthermore, accounting for the additional redshift evolution brings the gas fraction in the cluster outer radii into closer agreement with the universal baryon fraction.
\end{itemize}

Our overall conclusions are consistent with the view that the evolution of ICM thermodynamic profiles is influenced by baryonic physical processes, not only in the core regions but also extending beyond, into the bulk. These results highlight that the relative contribution of non-gravitational processes at different mass and radial scales in shaping the global thermodynamic properties of the intracluster gas remain poorly constrained. Further progress may be achieved by assessing the impact of cluster dynamical state using larger samples with robust morphological classifications, refining the modelling of intrinsic scatter with functional forms that better capture the shape of scaled thermodynamic profiles, and incorporating fully independent mass estimates such as those from weak lensing or galaxy velocity dispersions.

\begin{acknowledgements}
    EB, GWP, and EP acknowledge support from the Centre national d’études spatiales (CNES), France (ROR: https://ror.org/04h1h0y33) within the framework of the XMM-{\it Newton} mission.
\end{acknowledgements}

\bibliographystyle{aa} 
\bibliography{aa61466-26}

\begin{appendix}
\section{Mean matter density scaling}\label{app:mean_matt_dens}
An alternative choice for the background density used to calculate the cluster masses and to rescale the gas thermodynamic quantities is the mean matter density, $\rho_m(z)$. We defined the corresponding virial thermodynamic quantities with respect to an overdensity of $200$, so that the self-similar values of the thermodynamic quantities relative to those introduced in Section \ref{subsec:selfsimilar} with respect to $R_{500c},$ are given by
\begin{equation}
\label{eqn:rho500}\begin{split}
\rho_{200m} &=  200 \ \rho_m(z) = \rho_{500c} \ R(z), \\
T_{200m} &= T_{500c} \ \left(\frac{M_{200m}}{M_{500c}}\right)^{2/3} \ R(z)^{1/3}, \\
P_{200m} &= P_{500c} \ \left(\frac{M_{200m}}{M_{500c}}\right)^{2/3} \ R(z)^{4/3}, \\
K_{200m} &= K_{500c} \ \left(\frac{M_{200m}}{M_{500c}}\right)^{2/3} \ R(z)^{1/3}, \\
\end{split}
\end{equation}
where $R(z)=200 \cdot \Omega_m(z)/500$. Using the {\tt COLOSSUS} library \citep{colossus2018}, we computed the $R_{200m}$ and $M_{200m}$ for each cluster assuming an NFW profile with a concentration value equal to $c_{500c}=3.2$ \citep{pointecouteau2005}. 

\subsection{Application to the redshift scaling profiles}\label{app:sub:applic_alpha_z_r200m}
\begin{figure}
\centering
\includegraphics[width=8cm]{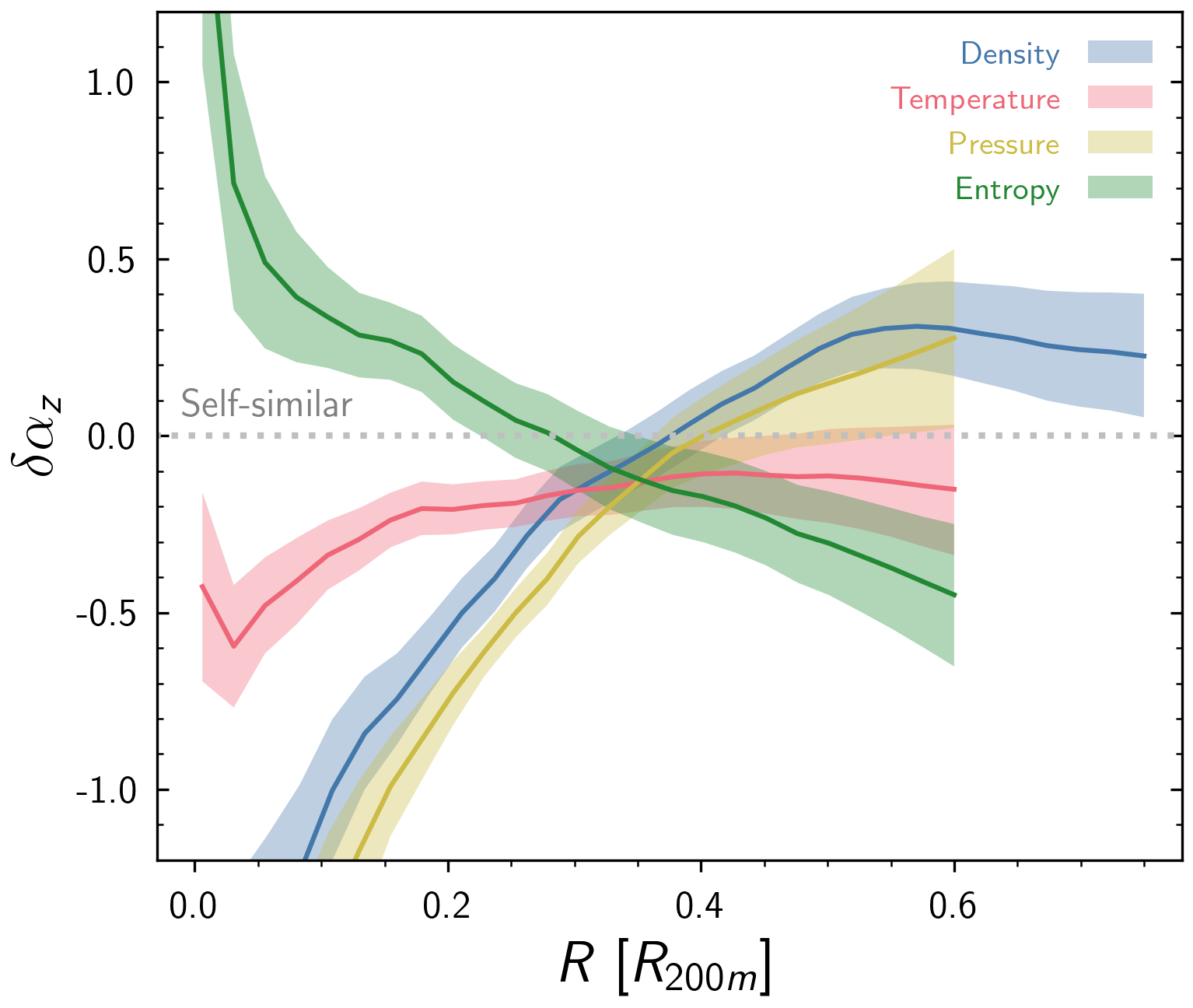}
\caption{\footnotesize Radial profiles of the redshift scaling factors relative to self-similar, obtained from density (blue), temperature (red), pressure (yellow) and entropy (green). The profiles are obtained following the piecewise power-law fitting procedure described in Sect.~\ref{subsec:modeling}, scaling all profiles for $M_{200m}$  and $R_{200m}$. The dotted grey line represents the expected values from a self-similar model. The shaded areas indicate the $1\sigma$ uncertainties.}
\label{fig:alpha_z:r200m}
\end{figure}

We rescaled our thermodynamic profiles by these quantities, and the radii to $R_{200m}$, and measured the $\delta\alpha_z$ and $\delta\alpha_M$ radial profiles following the procedure described in Section \ref{subsec:modeling}. The results for the $\delta\alpha_z$ profiles are shown in Figure \ref{fig:alpha_z:r200m}.
We do not report the results for the mass scaling profiles, as no significant differences are observed with respect to the results obtained using $R_{500c}$ scaling.
We observe that the redshift scaling profiles present strong deviations from self-similarity in the inner regions ($<0.4 \ R_{200m}$). We do not observe a measurable difference in the density redshift scaling profile at $\sim 0.7 \ R_{200m}\simeq 1.20 \ R_{500c}$; in contrast, the evolution of the temperature profiles is consistent with self-similar behaviour at $0.6 \ R_{200m}$, at the $2\sigma$ level. The redshift scaling factor profile of the pressure profiles indicates a stronger evolution than self-similarity at $\sim 0.6 \ R_{200m}$ which is in the opposite sense to that measured when the profiles are rescaled by $R_{500c}$, but with similar uncertainties, in agreement  with the predictions from \cite{nelson2014} and \cite{lau2015}. For entropy we observe a similar radial trend between the $R_{500c}$ and $R_{200m}$ scalings, with a very slightly increased normalisation. 

These results are in agreement with the predictions of \cite{nelson2014} and \cite{lau2015} for the inner regions, who found that the non-thermal pressure fraction increases when the profiles are rescaled by $R_{200m}$ (i.e. the pressure profiles are less self-similar in the inner regions). These authors also predict a reduction of the redshift dependence of the profiles in the outermost regions, at $R \gtrsim R_{200m}$. Unfortunately, our data do not probe the thermodynamic profiles beyond  $\sim 0.7 \ R_{200m}$, preventing a full test of this prediction.

\section{Extrapolations of thermodynamic profiles}\label{app:extrapol_prof}

To infer the mass and redshift scaling profiles relative to the self-similar predictions, we extended our analysis to the maximum reliable radius, extrapolating the observed profiles by accounting for both the statistical measurement uncertainties and the error propagation from $M_{500c}$, which directly impacts the scaled radial coordinate $x=R/R_{500c}$. For each profile, we undertook MCMC sampling of $10^4$ realisations of the radial values and the corresponding thermodynamic measurements. At each realisation, the sampled values were randomly drawn from a uniform distribution within their uncertainties. The sampled profiles were then extrapolated up to the desired radial limit using a linear B-spline interpolator. We derived the final extrapolated profiles and their associated uncertainties from the distribution of the realisations. The median was adopted as our best estimate, and the $16\%$ and $84\%$ percentiles defining the corresponding $1\sigma$ confidence intervals. 

\begin{figure*}
\centering
\includegraphics[width=16cm]{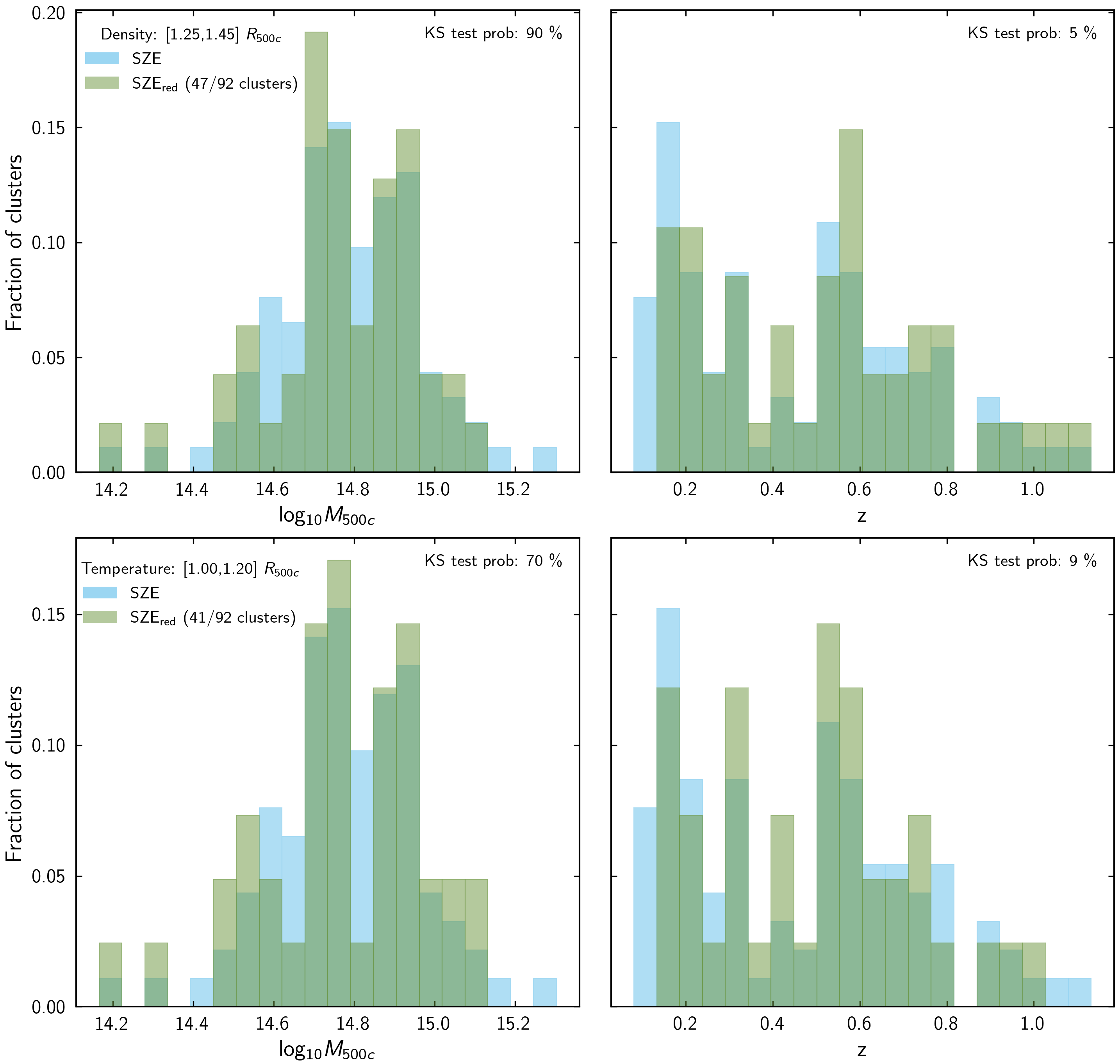} \\
\caption{\footnotesize Top panels: normalised histograms of the mass and redshift distributions of the SZE sample and the sample composed of clusters for which the density measurements are available in the $[1.25-1.45] \ R_{500c}$ radial bin (SZE$_{\rm red}$), in blue and green, respectively. We report the probability that the two sample are correlated using the KS test, as long as the number of clusters contained in the SZE$_{\rm red}$ sample. Bottom panels: same as the top panels for temperature measurements in the radial bin $[1.0-1.2] \ R_{500c}$.}
\label{fig:test_KS_dens_temp}
\end{figure*}

We investigated the robustness of extrapolation of the thermodynamic profiles beyond the radial limits by looking the representativeness of the sample composed of clusters for which measurements at large radii are available, namely SZE$_{\rm red}$, with respect to the entire sample of 92 galaxy clusters (SZE sample). The mass and redshift normalised distributions of the two samples for density measurements in the radial bin $R=[1.25, 1.45] \ R_{500c}$ are shown in the top panels of Fig.~ \ref{fig:test_KS_dens_temp} with blue and green histograms. Although the two samples present qualitatively the same mass distributions, confirmed by the high value of the Kolmogorov-Smirnov (KS) test probability, $90 \%$, the redshift distributions appear dissimilar and the KS test probability confirms this difference: $5\%$. Assuming a threshold of $15\%$ \citep[][]{dvoretzky_asymptotic_1956}, we assessed that the extrapolation of density profiles is robust up to $R=1.25\ R_{500c}$. 

A similar analysis of the mass and redshift distributions for temperature measurements in the radial bin $R=[1.0, 1.2] \ R_{500c}$ is presented in the bottom panels of Fig.~\ref{fig:test_KS_dens_temp}. The mass distributions are similar, with a KS test probability of $70\%$, but the redshift distributions have a KS test probability of $9\%$, moving it below the threshold. We then considered robust extrapolation up to $1.20$ $R_{500c}$ and $1.0$ $R_{500c}$ for density and temperature profiles, respectively. For pressure and entropy profiles we imposed the same upper limit radius of extrapolation as for the temperature profiles.

\section{Morphological classification} \label{app:morphological classification}
We subdivided the SZE sample, classifying the clusters according to their dynamical state, in order to infer the effects of cool cores and merger-related phenomena on the mass and redshift scaling profiles. The classification scheme is based on quantifying, in the log-log plane, the deviation of the scaled density ($n_{\textup{e}}/\langle n_{\textup{e}}\rangle_{R_{500c}}$) and scaled temperature\footnote{$T_{\textup{X}}$ is the spectroscopic temperature measured in the $[0.15-0.75]$ $R_{500c}$} ($T/T_{\textup{X}}$) profiles from the median scaled profiles.
The density profiles were normalised to the mean density within $R_{500c}$, so the dispersion reflects only variations in shape. Comparison with \cite{arnaud_universal_2010} shows that this classification scheme broadly reproduces the cuts in $\langle w \rangle$ that were applied to \rexcess.

We evaluated the deviations in density and temperature within two radial bins, $[0.01, \ 0.5] \ R_{500c}$, with respect to the median scaled profiles. Where the thermodynamic profiles were not defined we proceeded to extrapolate them at specific radial bins (see Appendix \ref{app:extrapol_prof}). Figure~\ref{fig:CoolCore} shows the anti-correlation of the density and temperature deviations evaluated at $R=0.01 \ R_{500c}$ for the SZE-selected sample (filled circles). The \rexcess \ sample classification (indicated by empty markers) was performed in \cite{pratt_galaxy_2009} by computing the central density and the cooling time \citep[][]{ohara_effects_2006, croston_galaxy-cluster_2008}. A Spearman rank test found a probability of $10^{-6}$ that the anti-correlation occurs by chance for the SZE-selected sample. The correlation disappears at large radii, finding a probability of $0.4$ at $R=0.5 \ R_{500c}$. We classified as cool core (CC) clusters those systems with $\Delta [\textup{log}_{10} (T)]<-0.03$ and $\Delta [\textup{log}_{10} (n_{\textup{e}})]>0.03$, and as morphological disturbed clusters those systems with $\Delta [\textup{log}_{10} (T)]>0.03$ and $\Delta [\textup{log}_{10} (n_{\textup{e}})]<-0.03$. These thresholds, defined to select the extremes of the distribution, can be applied equally to simulations.

\begin{figure*}
\centering
\includegraphics[width=16cm]{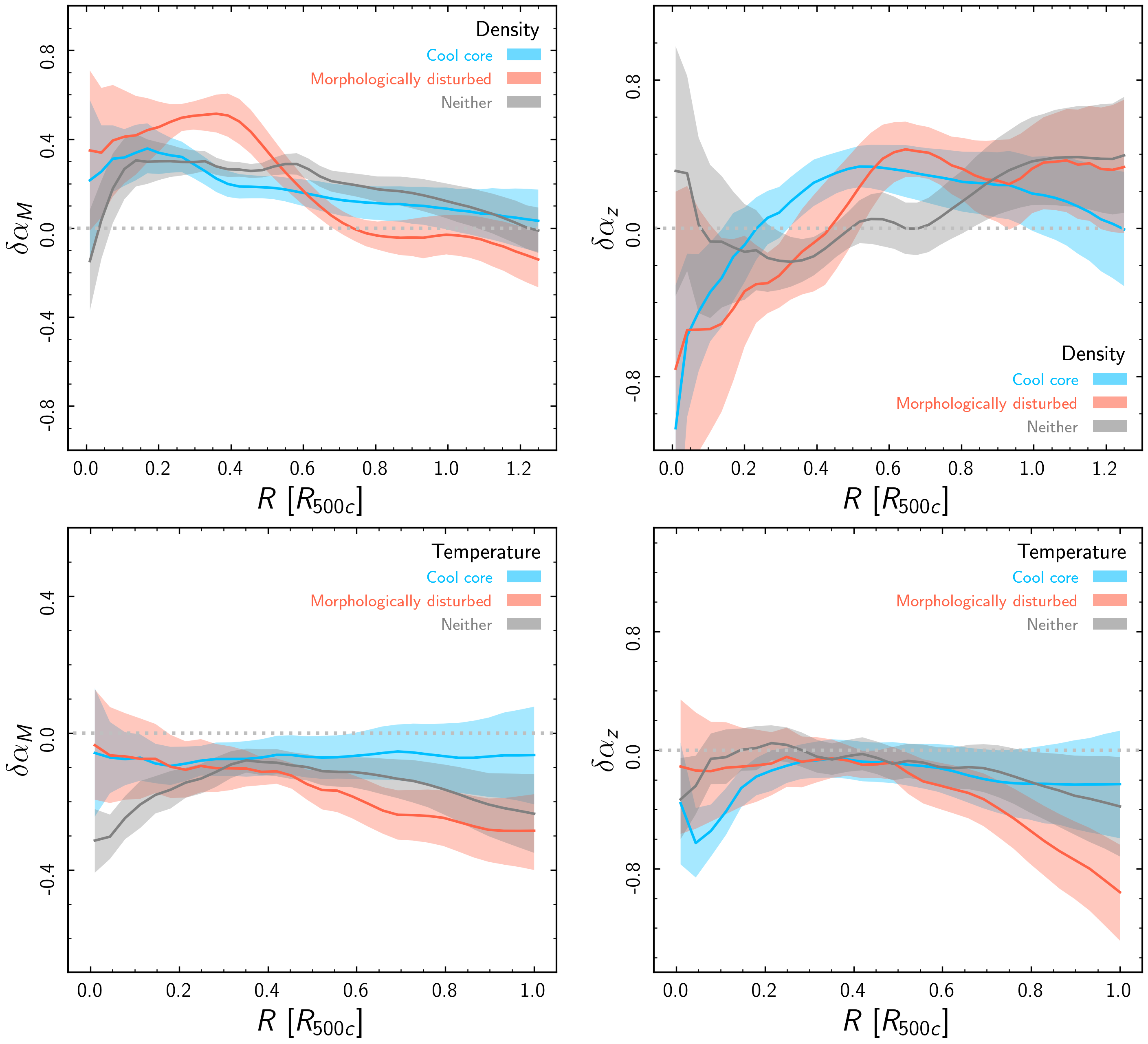} 
\caption{\footnotesize Top panels: radial profiles of the mass (left-hand panel) and evolution (right-hand panel) scaling for density for the classified relaxed clusters (cool core, in cyan), morphological disturbed clusters (in red), and the clusters not classified as relaxed or disturbed (in grey). The dashed grey line indicates the predicted self-similar value. Bottom panels: corresponding radial scaling profiles of the mass (left-hand panel) and evolution (right-hand panel) for temperature profiles.}
\label{fig:dens_temp_CC_vs_NCC_plots}
\end{figure*}

We applied the fitting procedure described in Sect.~\ref{sec:fitting_procedure} to the three subsamples defined from the density and temperature profiles. Figure \ref{fig:dens_temp_CC_vs_NCC_plots} shows the radial variation of the mass and redshift scaling factors, relative to self-similar, for our three cluster subsamples: cool cores (26 objects, cyan), morphologically disturbed (23 objects, red), and clusters classified as neither cool core nor morphologically disturbed (43 objects, black), for density (top) and temperature (bottom). We find no significant differences in the scalings of density or temperature across these populations. The limited sample size in each subsample currently precludes more robust constraints on the radial dependence of these scaling factors.

\begin{figure}
\centering
\includegraphics[width=8cm]{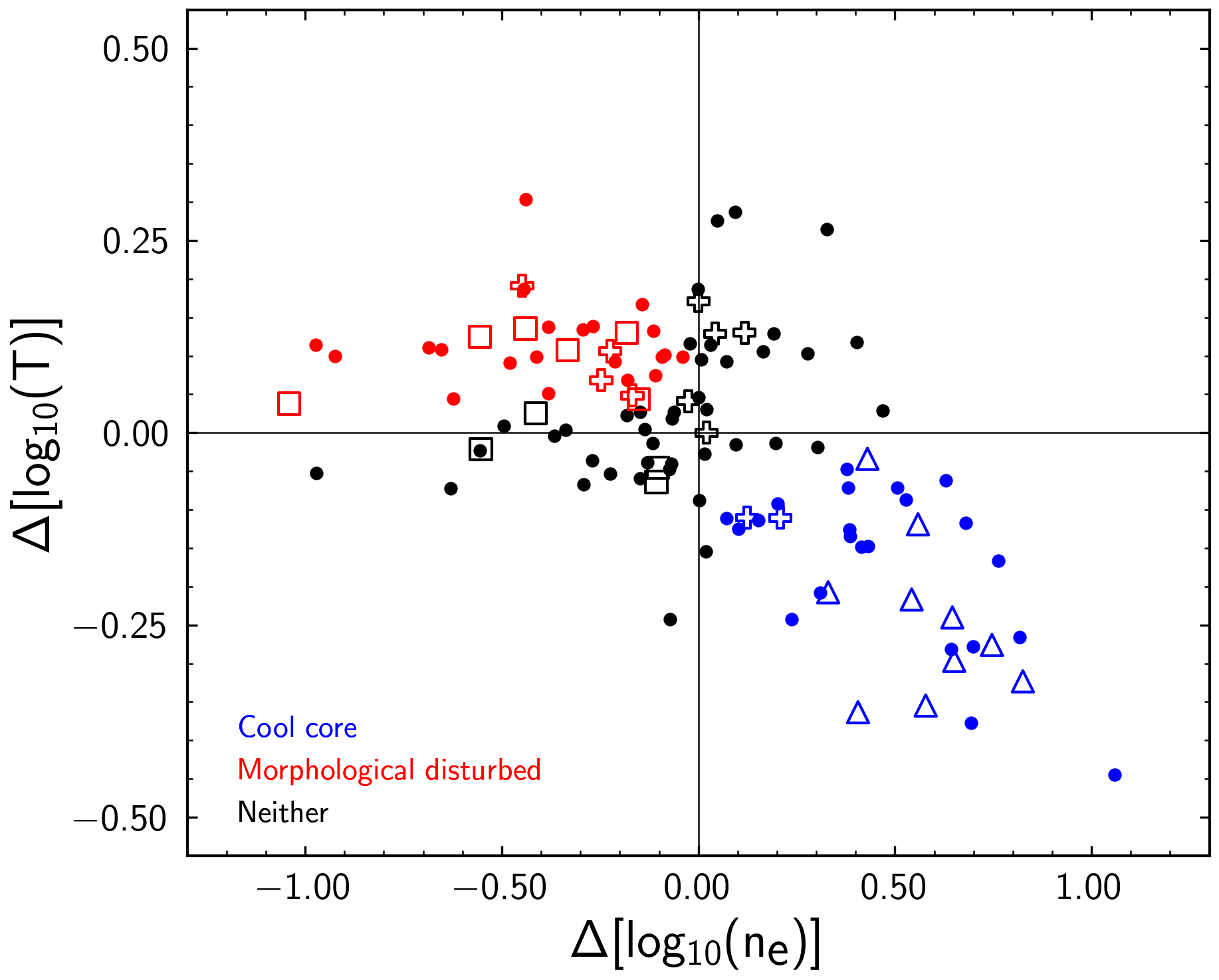}
\caption{\footnotesize Logarithmic scaled density deviation from the median profile versus the corresponding logarithmic scaled temperature deviation computed at $R=0.01 \ R_{500c}$. Data corresponding to the SZE-selected and \rexcess \ samples are marked with filled and open markers, respectively. Each system is colour-coded according to its dynamical state: cool core clusters (blue), morphological disturbed clusters (red), and clusters classified as neither relaxed nor disturbed (black). The clusters in \rexcess \ are colour-coded according to our morphological classification but with different marker shapes according to \cite{pratt_galaxy_2009} classification. The squares, triangles and pluses represent cool core, disturbed, and neither relaxed nor disturbed clusters, respectively.}
\label{fig:CoolCore}
\end{figure}

\section{Marginalised posterior likelihood}\label{app:marg_post_likeli}
\begin{figure*}
\centering
\resizebox{\hsize}{!}{\includegraphics{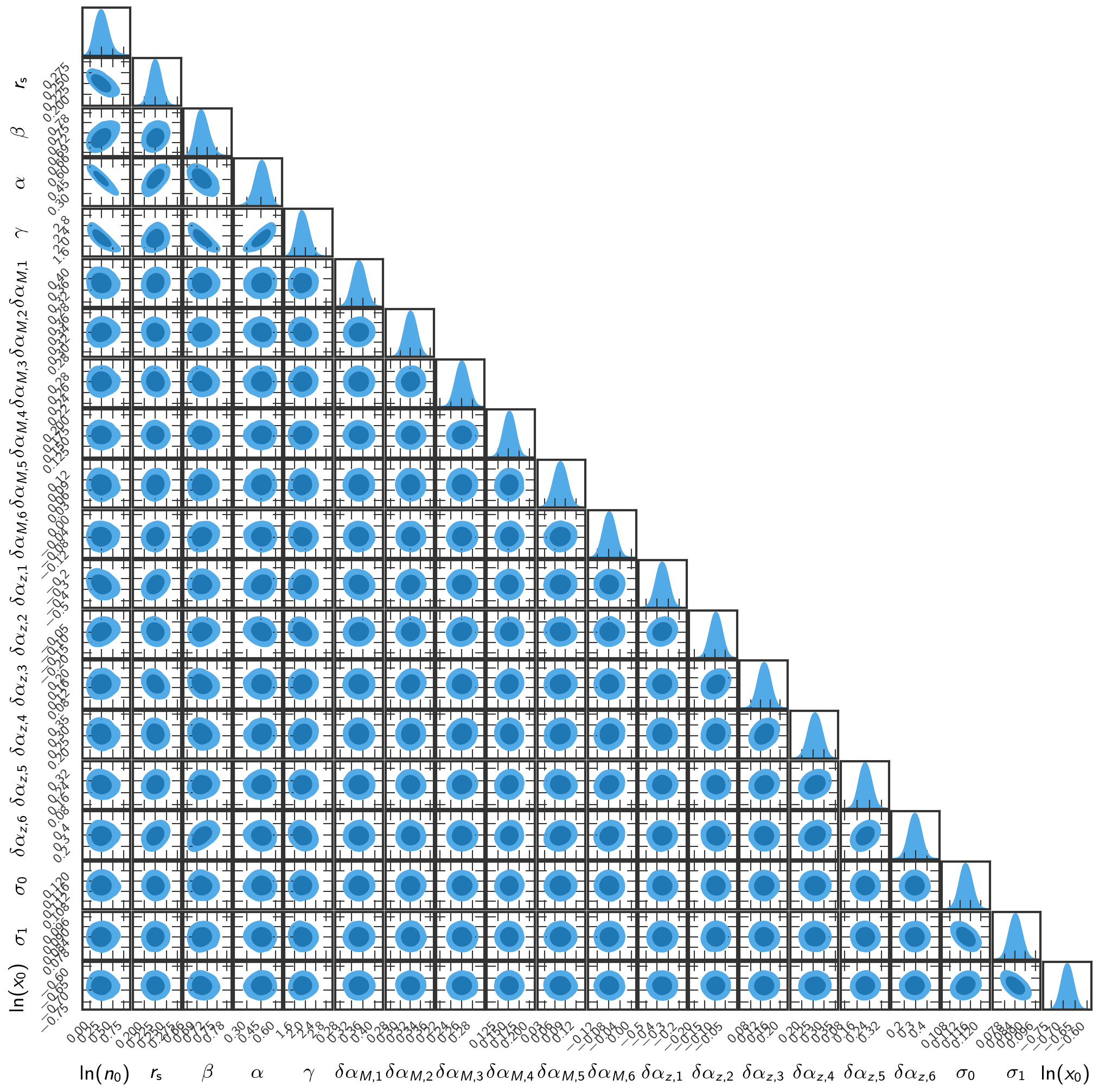}} 
\caption{\footnotesize Marginalised posterior likelihood for the parameters of the best-fitting density profile model, as detailed in Section \ref{subsec:densmodel}. The priors of these parameters are listed in Table ~\ref{table:prior_and_best_fit}.}
\label{fig:cornerplot_density}
\end{figure*}

\begin{figure*}
\centering
\resizebox{\hsize}{!}{\includegraphics{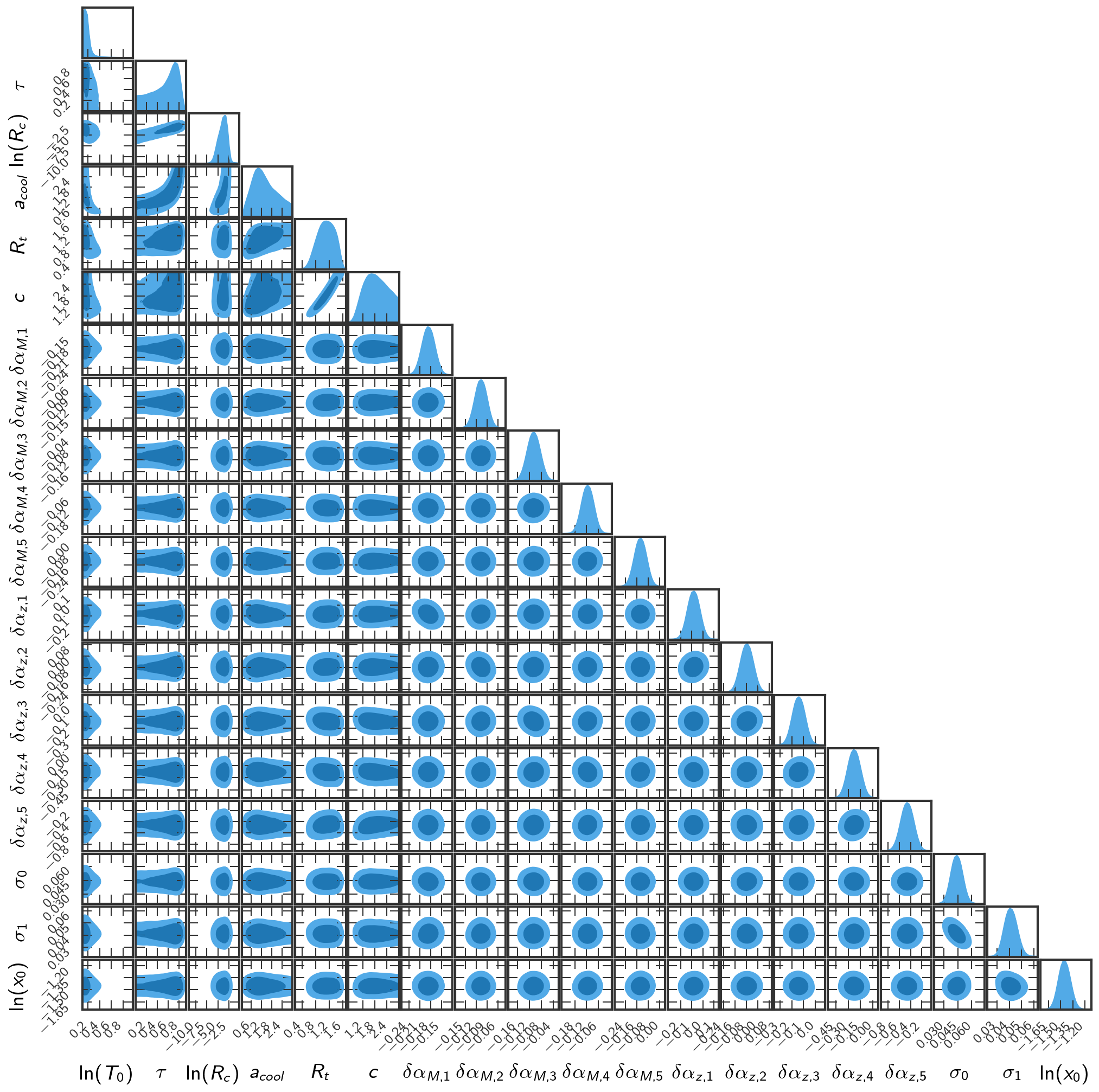}} 
\caption{\footnotesize Marginalised posterior likelihood for the parameters of the best-fitting temperature profile model, as detailed in Section \ref{subsec:tempmodel}. The priors of these parameters are listed in Table ~\ref{table:prior_and_best_fit}.}
\label{fig:cornerplot_temperature}
\end{figure*}

Figure \ref{fig:cornerplot_density} and \ref{fig:cornerplot_temperature} show the 1D and 2D posterior distributions for the parameters derived from the density and temperature profile fits in Section \ref{subsec:thermod_profiles_results}.

\end{appendix}

\raggedright
\end{document}